\documentclass[a4paper,11pt]{article}
\pdfoutput=1 % if your are submitting a pdflatex (i.e. if you have
\usepackage{jheppub} % for details on the use of the package, please
\usepackage[T1]{fontenc} % if needed

\newcommand{\be}{\begin{eqnarray}}
\newcommand{\ee}{\end{eqnarray}}
\newcommand{\kt}{\tilde{\kappa}}
\newcommand{\kap}{\kappa}

\title{\boldmath Non-perturbatively gauge-fixed compact $U(1)$ lattice gauge theory}

\author{Asit K. De}
\author{and Mugdha Sarkar}
\affiliation{Theory Division, Saha Institute of Nuclear Physics, HBNI,\\1/AF Salt Lake, Kolkata 700064, India}

\emailAdd{asitk.de@saha.ac.in}
\emailAdd{mugdha.sarkar@saha.ac.in}
\abstract{An extensive study of the compact $U(1)$ lattice gauge theory with a higher derivative gauge-fixing term and a suitable counter-term has been undertaken to determine the nature of the possible continuum limits for a wide range of the parameters, especially at strong gauge couplings ($g>1$), adding to our previous study at a single gauge coupling $g=1.3$ \cite{DeSarkar2016}. Our major conclusion is that a continuum limit of free massless photons (with the redundant pure gauge degrees of freedom decoupled) is achieved at any gauge coupling, not necessarily small, provided the coefficient $\kt$ of the gauge-fixing term is sufficiently large. In fact, the region of continuous phase transition leading to the above physics in the strong gauge coupling region is found to be analytically connected to the point $g=0$ and $\kt \rightarrow \infty$ where the classical action has a global unique minimum, around which weak coupling perturbation theory in bare parameters is defined, controlling the physics of the whole region. A second major conclusion is that, local algorithms like Multihit Metropolis fail to produce faithful field configurations with large values of the coefficient $\kt$ of the higher derivative gauge-fixing term and at large lattice volumes. A global algorithm like Hybrid Monte Carlo, although at times slow to move out of metastabilities, generally is able to produce faithful configurations and has been used extensively in the current study.}

\begin{document} 
\maketitle
\flushbottom

\section{Introduction}\label{Intro}
The Fadeev-Popov gauge-fixing scheme and the resulting Becchi-Rouet-Stora-Tyutin (BRST) symmetry has been of great importance in perturbative definitions of gauge theories. However, this scheme is not directly applicable to compact gauge fields used in euclidean lattice regulators for non-perturbatively defined gauge theories. Expectation value of a gauge-invariant operator in a BRST-invariant theory with compact gauge fields returns an indeterminate $0/0$ value \cite{Neub87,Testa98}; the real-valued ghost field determinant changes sign due to presence of Gribov copies, and the BRST symmetry ensures an exact cancellation of contributions with opposite signs. The proposal for a remedy evading the above no-go situation, in case of non-abelian gauge theory, is the so-called equivariant BRST (eBRST) scheme that gauge-fixes only the coset space, leaving a nontrivial Cartan subgroup invariant \cite{Schad99, GoltSham2004}. For the abelian case, a nonperturbative gauge-fixing scheme as proposed by \cite{GoltSham97, Sham98} includes a specially engineered higher derivative (HD) gauge-fixing term in the lattice action breaking BRST invariance. Recovery of gauge symmetry in the physical sector is to be achieved by tuning appropriate counter-terms.      

For Yang-Mills theories on a discrete space-time euclidean lattice, Wilson in his seminal paper \cite{Wilson74} introduced a manifestly gauge-invariant formalism, that works equally well for abelian theories, in terms of a functional integral with a gauge-invariant Haar measure using group-valued fields on the links of the lattice (so that the algebra-valued gauge fields are compact). For vector-like theories like QCD and QED, where gauge-invariance can be maintained on the regulator, Wilson's gauge-invariant scheme is highly successful, although gauge-fixing may be necessary for matching to a perturbative renormalisation scheme like $\overline{MS}$. However, given that fermions on the lattice needs to explicitly break chiral symmetry \cite{KarstenSmit81, NN81a, NN81b, GinspargWilson82}, for chiral gauge theories gauge-invariance is broken on the lattice regulator\footnote{
In a different approach to lattice chiral gauge theory that modifies chiral symmetry on lattice according to the Ginsparg-Wilson relation \cite{GinspargWilson82}, one arrives at a link field dependent fermion measure and an exact solution to the integrability condition on the space of the lattice link fields was obtained in the Abelian case \cite{Luscher98}.
}. In a manifestly local formalism of lattice chiral gauge theories, the standard Wilson method with a gauge-invariant measure leads to various problems \cite{BockDeSmit91, GoltPetchSmit91}. The problems arise due to the undesired presence, in such theories, of the longitudinal gauge degrees of freedom ({\it lgdof}) which become manifest as physical degrees of freedom (as scalar fields) in a gauge-noninvariant theory such as manifestly local proposals of lattice chiral gauge theory \cite{Smit, Swift, EP, DWW}. The interactions of the {\it lgdof} with the physical sector can generally be strong since any point on the gauge orbit is as likely as any other in a gauge-invariant formalism. The usual method to tackle such situations is to give the {\it lgdof} a dynamics through a particular gauge-fixing mechanism that is expected to control or tame the {\it lgdof}. A general gauge-fixing scheme for compact gauge fields associated with the lattice link fields, applicable at all strengths of the interaction including nonperturbative values, is thus very welcome. In addition, a BRST-like general framework has long been considered a satisfactory way to define a gauge theory.

In this paper, continuing our work from before \cite{BasakDeSinha2004, DeSarkar2016}, we present results from our numerical simulations of nonperturbative gauge-fixing for the abelian case as proposed in \cite{GoltSham97, Sham98}. Our work on the non-abelian case involving eBRST formalism is in progress and will be reported elsewhere.

The HD gauge-fixing proposal for the abelian gauge theory was studied extensively in the weak gauge coupling region some time ago \cite{Bock_etal2000} with results as anticipated in the proposal. For sufficiently large coefficient (denoted as $\kt$ in this paper) of the HD gauge-fixing term, there is a novel \emph{continuous} phase transition (called the FM-FMD\footnote
{FM and FMD stand respectively for ferromagnetic and ferromagnetic directional phases. The nomenclature is derived from the phases of the theory in the so-called reduced limit, i.e., when the gauge coupling tends to zero, leaving the theory to be entirely a HD scalar theory.}
transition in the literature, FMD being a spatially modulated ordered phase, a novel phase with broken Euclidean symmetry). When this transition is approached from the FM-side (regular ordered phase that respects Euclidean symmetry), the spectrum contains only free massless photons and the scalar fields ({\it lgdof}) decouple. These results have been explicitly verified in weak gauge coupling region using both perturbative analysis and through Monte Carlo importance sampling by numerical simulation.  

Strong coupling of the {\it lgdof} with chiral fermions is what led to the failure of a prevalent class of non-perturbative chiral gauge theory proposal \cite{BockDeSmit91, GoltPetchSmit91}. With the success of the HD gauge-fixing model of the compact $U(1)$ lattice gauge theory in decoupling the {\it lgdof}, feasibility of manifestly local abelian chiral gauge theories on lattice was shown for Wilson fermions \cite{Bock_etal1998, Bock_etal1998_2} and also for lattice domain wall fermions \cite{BasakDe2001, BasakDe2001_2}. It is worth mentioning here that, in the standard Wilsonian definition of a lattice gauge theory (that is, without gauge-fixing), the strong coupling of the unphysical {\it lgdof} with fermions (or with any physical degrees of freedom) in a gauge-noninvariant theory like a lattice chiral gauge theory, is irrespective of the strength of the usual gauge coupling. In fact, almost all studies in this area have been done in the so-called reduced model (i.e., in the limit of gauge coupling going to zero) and the basic problems are already present there.   

All the success of the HD gauge-fixing approach for the abelian theory is, so far, mainly in the reduced limit or in the weak gauge coupling region. The question naturally arises in a general framework as to what happens when the bare gauge coupling is not necessarily small. This question gains even more importance with the arrival of the eBRST gauge-fixing proposal \cite{GoltSham97, Sham98} for the manifestly local non-abelian chiral gauge theory on the lattice, since after partial gauge-fixing in the coset space $\grave{a}\;\; la$ eBRST, finally an abelian part will remain unfixed. If this is left unfixed, the {\it lgdof} will again surface and the chiral theory will be spoiled, as described above briefly. Hence the machinery of the HD gauge-fixing for the remaining abelian part will again be necessary and, therefore, a comprehensive understanding of the HD gauge-fixing scheme at a broad range of the gauge coupling is desirable.      

A first preliminary account in this direction was presented, some time ago, in \cite{BasakDeSinha2004}. The novel FM-FMD transition, that was responsible for the decoupling of the {\it lgdof} and the emergence of the original gauge symmetry, was still found to be present at stronger gauge couplings, with bare values larger than unity. With large gauge couplings, the FM-FMD transition was first order for small values of $\tilde{\kappa}$ (coefficient of the HD gauge-fixing term). Only at large $\tilde{\kappa}$, the transition was  found to be continuous, with a tricritical point separating it from the first order transition. However, the nature of the possible continuum limit while approaching
 the continuous part of the transition from the FM-side was not studied. As a result the emerging physics at this transition was not clear.        

Only recently, as mentioned in \cite{DeSarkar2016}, we have realised the limitations of the configuration-generating algorithm, multihit Metropolis (MM), used in all earlier studies in the weak gauge coupling region \cite{Bock_etal2000} and also in the earlier work at strong coupling \cite{BasakDeSinha2004}. As we shall see in the next Section, the HD gauge-fixing term involves the square of the gauge-covariant lattice laplacian, and as a result the action density at a lattice site involves significantly more lattice links than the usual Wilson plaquette. For a local updating algorithm like the MM, the accept/reject step is incorporated after each local update of the field. As a result, unless the coefficient $\kt$ of the HD-term is sufficiently small, the algorithm would struggle to generate legitimate field-configurations. With stronger gauge couplings, the physically interesting continuous FM-FMD phase transition takes place only at larger values of $\kt$, and this leads to the failure of the MM algorithm. In this paper, we shall give a more detailed account of why we needed to abandon MM, irrespective of the number of hits and present a comparison with the Hybrid Monte Carlo (HMC), a global algorithm, generally adopted for generation of field configurations in this paper.

In \cite{DeSarkar2016}, results of the phase diagram and the emerging physics at the continuous part of the FM-FMD transition obtained with the newly employed HMC algorithm for the HD action were presented only at one value of the strong gauge coupling, viz., $g=1.3$. The current work aims at a consistent and comprehensive picture applicable in general for a wide range of parameters to determine especially how the strong gauge coupling phase diagram gradually emerges from that of the weak gauge coupling region and how the two regions are related, if at all.  

The main result of this paper is that the Lorentz covariant physics emerging in the strong gauge coupling region, by approaching the FM-FMD transition from the FM-side, is actually governed by that at the phase transition at $g=0$, $\kt \rightarrow \infty$ and $\kap \sim 0$, $\kap$ being the coefficient of a dimension-2 mass counter-term required to recover gauge symmetry (In the weak gauge coupling limit $g=0$, the action has a unique global minimum). The paper also establishes the inadequacy of a local algorithm like MM for larger $\kt$ and bigger volumes, by comparing results with the HMC algorithm at different regions of the coupling parameter space. As remarked above, a tricritical line emerges for $g>1$ in the 3-dimensional phase diagram separating the first order and the continuous FM-FMD transition surfaces. A detailed investigation of the universality class of the tricritical line is outside the scope of the current work.

The paper is organised as follows.  The next section, i.e., Sec.\ref{Review} presents a review of the main ideas of the HD gauge fixing action for the compact $U(1)$ lattice gauge theory based on the theory at weak gauge coupling. After briefly describing, in Sec.~\ref{HMC}, the implementation of the force terms during the molecular dynamics trajectory of the HMC algorithm applied to the gauge-fixing theory with a HD action, and definitions of all quantities measured, we present a comparison in Sec.~\ref{Algorithms} of the MM and the HMC algorithms. In Sec.~\ref{Phases}, we present results of the phase diagram at several values of the gauge coupling $g$ including at the end a schematic 3-dimensional phase diagram covering a wide range of parameters from the weak gauge coupling to the strong gauge coupling regions. We collect results of various two point functions, and also the average plaquette value in Sec.~\ref{Measurements} to understand the physics of the FM phase while approaching the FM-FMD transition at strong gauge couplings. We present our main conclusions in Sec.~\ref{Conclusion}.

\section{The Abelian Gauge Fixing Theory on Lattice} \label{Review}
In this section, we briefly review the compact $U(1)$ gauge theory with the HD gauge-fixing term and mention its salient features validated mostly through analytic and numerical investigations, done earlier, at weak gauge couplings. Detailed accounts are found in \cite{GoltSham97, Bock_etal2000, Bock1997fu}.
 
The euclidean action on a 4-dimensional hypercubic lattice is given by: 
\be
S = S_{\rm W} + S_{\rm GS} +  S_{\rm ct}.
\label{S}
\ee 

As we shall see in the following, the action $S$ explicitly contains only physical fields and no ghost fields\footnote
{The compact lattice $U(1)$ gauge fields are self-interacting and in principle the action could include ghost fields which would then be expected to decouple only in the continuum limit in the standard scenario. 
}. 
The gauge symmetry of the first term $S_{\rm W}$ is explicitly broken by the gauge-fixing second term $S_{\rm GS}$ and also by the third term $S_{\rm ct}$ in the above action.

The first term in (\ref{S}), $S_{\rm W}$, is the gauge-invariant standard Wilson term containing a summation over all gauge plaquettes $U_{{\rm P}\mu\nu}(x)$,
\be \label{W}
S_{\rm W} = \frac{1}{g^2}\sum_{x, \;\mu<\nu} \left( 1 - {\rm Re} \, U_{{\rm P}\mu\nu}(x) \right ),
\ee
the plaquette being the smallest Wilson loop around an elementary square at a lattice point $x$ on the $(\mu,\nu)$ plane.

The second term in (\ref{S}), $S_{\rm GS}$, is the Golterman-Shamir HD gauge-fixing term \cite{GoltSham97, Sham98} and is given by
\be \label{GS}
S_{\rm GS}={\kt}\left ( \sum_{xyz} \Box_{xy}(U)\, \Box_{yz}(U) - \sum_x  B_x^2 \right ),
\ee
where the gauge-covariant Laplacian $\Box_{xy}(U)$ is given by, 
\be
\Box_{xy}(U) = \sum_\mu ( \delta_{y,x+\mu} U_{x\mu} + \delta_{y,x-\mu}U^\dagger_{x-\mu,\mu} - 2 \delta_{xy} ), 
\ee
and,
\be
B_x = \sum_\mu ({\cal{A}}_{x-\mu,\mu} + {\cal{A}}_{x\mu})^2/4, \;
{\rm with} \; {\cal{A}}_{x\mu}={\rm Im} U_{x\mu}.
\ee

The third term in (\ref{S}), $S_{\rm ct}$, generally represents a collection of all possible counter-terms, needed to ensure recovery of gauge symmetry at a \emph{desirable} continuous phase transition. The counter-terms are determined by usual power counting which is validated by the choice of a gauge in $S_{\rm GS}$ that is expected to be the renormalisable Lorentz covariant gauge in the continuum. In principle, $S_{\rm ct}$ contains a dimension-2 gauge field mass counter-term, and five marginal counter-terms, allowed by the exact lattice symmetries \cite{Borelli_etal1990}. Three of the five marginal counter-terms are field renormalisation counter-terms for the gauge field, and the other two counter-terms are to nullify quartic gauge field self-interaction. It has been argued in \cite{Bock_etal2000} that all the marginal counter-terms in this theory can be perturbatively treated. However, being perturbative, they cannot give rise to a new phase transition. We consider, 
\be \label{ct}
S_{\rm ct} = -\kap \sum_{x\,\mu} \left ( U_{x\mu} + U^\dagger_{x\, \mu}  \right ),
\ee
which is a dimension-2 mass counter-term, as apparent from expanding the lattice gauge field $U_{x\mu} = \exp(iagA_\mu(x))$ for small lattice spacing $a$. As we shall witness later, the dimension-2 mass counter-term is the one responsible for the FM-FMD phase transition giving rise to a new universality class near that transition. 

It can be explicitly shown \cite{GoltSham97} that the action (\ref{S}) with the HD gauge-fixing term has a unique absolute minimum at $U_{x\mu} =1$. In the naive continuum limit (i.e., lattice spacing $a \searrow 0$ in the action), the HD gauge-fixing term becomes the familiar covariant gauge fixing term 
\be \label{cov-gf}
\kt g^2 \int d^4x (\partial_\mu A_\mu)^2 = (1/2\xi)\int d^4x (\partial_\mu A_\mu)^2 \label{cov_gf},
\ee
where $\xi$ is defined as  
\be
\xi = 1/(2\kt g^2).   \label{xi}
\ee
The above considerations validate a weak coupling perturbation theory (WCPT) of the gauge fixed theory with $\xi\sim 1$ around $g=0$ and large $\kt \rightarrow \infty$.

From Eq.~(\ref{xi}), it is clear that, to keep $\kt g^2$ or $\xi$ of ${\cal O}(1)$, we need to tune $\kt\nearrow \infty$ as the gauge coupling $g\searrow 0$. In practice, for a given gauge coupling $g$, it needs to be seen how large the coefficient $\kt$ of the HD gauge-fixing term needs to be in order for the gauge-fixing term to take discernible effect. It can be expected that for weak gauge couplings, there would be no significant effect of gauge-fixing for very small values of $\kt$, since the effective coefficient of the gauge-fixing term given in Eq.~(\ref{cov_gf}) is then really tiny. With increase of the value of $\kt$, but still with weak gauge couplings, gauge fixing can be expected to take effect, as has been found in investigations. However, what happens at strong gauge couplings cannot be guessed at all and is the theme of ref.~\cite{DeSarkar2016} and the current investigation. Numerical simulations can find out how large $\kt$ needs to be for a given $g$.  

The theory is defined by the following functional integral for the partition function,
\be
Z = \int {\cal D}U \exp(-S[U_{x\mu}]), \label{Z}
\ee 
with $S[U_{x\mu}]$ given by (\ref{S}) and
\be \label{Haar}
{\cal D}U = \prod_{x\mu} dU_{x\mu}, 
\ee
where $dU_{x\mu}$ is the gauge invariant Haar measure.

Writing the gauge non-invariant part of the action (\ref{S}) collectively as
\be
S_{\rm NI}[U_{x\mu}] = S_{\rm ct}[U_{x\mu}]+S_{{\rm GS}}[U_{x\mu}], 
\ee
let us consider a gauge transformation $U_{x\mu} \rightarrow g_x U_{x\mu} g^\dagger_{x+\mu}$ ($g_x\in U(1)$) in the partition function (\ref{Z}) (remembering that ${\cal D}U$ and $S_{\rm W}$ are gauge-invariant while $S_{\rm NI}[U_{x\mu}]$ is not),
\be
Z &=& \int {\cal D}U \exp \left(-S_{\rm W} - S_{\rm NI}[U_{x\mu}]\right ) \\
&\rightarrow & \int {\cal D}U \exp \left(-S_{\rm W} - S_{\rm NI}[g_x U_{x\mu}g^\dagger_{x+\mu}]\right ) \\
& = & \int {\cal D}g {\cal D}U \exp \left(-S_{\rm W} - S_{\rm NI}[g_x U_{x\mu}g^\dagger_{x+\mu}]\right ) \\
& = & \int {\cal D}\phi {\cal D}U \exp \left(-S_{\rm W} - S_{\rm NI}[\phi^\dagger_x U_{x\mu}\phi_{x+\mu}]\right ),
\ee
where in the penultimate step, we multiply each side by $\int {\cal D}g = \prod_x \int dg_x = 1$ (normalised gauge volume at each site), and in the final step, $\phi_x \equiv g^\dagger_x$ has been used.  

As is apparent from the above steps, under a gauge transformation $U_{x\mu} \rightarrow g_xU_{x\mu}g^\dagger_{x+\mu}$, the gauge non-invariant terms pick up the {\it lgdof}, and the theory becomes a scalar-gauge system with $S_{\rm NI}[\phi^\dagger_xU_{x\mu}\phi_{x+\mu}]$.   

The action obtained after the gauge transformation (the so-called Higgs picture) involves both the gauge fields and the {\it lgdof} which are essentially radially frozen scalar fields $\phi_x$. 

The mass counter-term (\ref{ct}) takes the following form in the Higgs picture:
\be
S^\phi_{\rm ct} = -\kap \sum_{x\,\mu} \left ( \phi_x^\dagger U_{x\mu}\phi_{x+\mu} + \phi_{x+\mu}^\dagger U^\dagger_{x\, \mu}\phi_x  \right )
\sim -\kap \sum \phi^\dagger \Box(U) \phi,
\ee 
which is the usual kinetic term for the scalar field.

Similarly, the HD gauge-fixing term (\ref{GS}) becomes, in the Higgs picture,
\be
S^\phi_{\rm GS}= \kt \left ( \sum  \phi^\dagger \Box^2 (U) \phi - \sum {\mathcal{B}}^2 \right ),
\ee
where,
\be \label{B-Higgs}
{\mathcal{B}}_x = \sum_\mu (\bar{\cal{A}}_{x-\mu,\mu} + \bar{\cal{A}}_{x\mu})^2/4, \;
{\rm with} \; \bar{\cal{A}}_{x\mu}={\rm Im} \left ( \phi^\dagger_x U_{x\mu}\phi_{x+\mu} \right ).
\ee

The total action, in the Higgs picture, thus assumes the form:
\be \label{S-Higgs}
S^\phi = S_{\rm W} + S^\phi_{\rm GS} +  S^\phi_{\rm ct}
\ee
where the standard Wilson term $S_{\rm W}$ is gauge invariant and hence does not pick up the {\it lgdof} when the functional integral integrates along the gauge orbit.

The gauge invariance as found in the standard Wilson term $S_{\rm W}$ alone is the target symmetry under the gauge transformations:
\be \label{g-sym}
U_{x\mu} \rightarrow g_xU_{x\mu}g^\dagger_{x+\mu},\;\;\; g_x\in U(1)
\ee

However, the total action (\ref{S-Higgs}) in the Higgs picture has enlarged, unphysical symmetry under the transformations 
\be \label{h-sym}
U_{x\mu} \rightarrow h_xU_{x\mu}h^\dagger_{x+\mu},\;\;\; \phi_x \rightarrow h_x\phi_x,\;\;\;
h_x\in U(1).
\ee
We would call the local symmetries given by (\ref{g-sym})  and (\ref{h-sym}) respectively as the $g$-symmetry (target physical symmetry) and the $h$-symmetry.
 
Putting $\phi_x =1$ in the expression for the action $S^\phi$ in the Higgs picture (\ref{S-Higgs}) recovers the action (\ref{S}), called the action in the vector picture. Given the Haar measure (\ref{Haar}) of the functional integrals, theories given by the two actions (\ref{S}) and (\ref{S-Higgs}) are completely equivalent. 

With vanishing $\kt$, the theory approaches an abelian gauge-Higgs system.

With zero gauge coupling $g=0$, we have $U_{x\mu}=1$ for all the links of the lattice. This is known as the reduced limit. The reduced model is defined by the functional integral,
\be 
Z_{\mathrm{red}} = \int {\cal D}\phi \, \exp{(-S[\phi])},
\ee 
with,
\be \label{S-reduced}
S[\phi] = -\kap \sum_x \phi^\dagger_x \left (\Box\phi \right )_x + \kt \sum_x \left \{\phi^\dagger_x\left ( \Box^2 \phi \right )_x - b_x^2 \right \},
\ee
where $b_x$ is the appropriate modification of ${\mathcal{B}}_x$ of Eq.~(\ref{B-Higgs}) with $U_{x\mu}=1$.

The reduced model action is invariant under the global transformations
\be \label{red-sym}
\phi_x \rightarrow h \, \phi_x ,
\ee
where $h\in U(1)_{\mathrm{global}}$ is independent of the lattice site. 

At $\kt=0$, the reduced model is just the radially frozen scalar field theory in 4 dimensions with $U(1)$ global symmetry. This is also known as the XY model, or as the non-linear sigma model with global $U(1)$ symmetry, in 4 dimensions. The phase diagram of this theory is well known. At large $\kap$, the system is frozen, i.e., $|\langle \phi_x \rangle | = 1$ with perfect ferromagnetic (FM) ordering. As $\kap$ is lowered, due to quantum fluctuations, there is a continuous phase transition of the system at $\kap = \kap_{\rm FM-PM} = 0.15$ (numerically determined) into a paramagnetic (PM) phase where $|\langle \phi_x \rangle | = 0$.  Because of the symmetry under $\kap \rightarrow -\kap$ and $\phi_x \rightarrow \phi_x^{\rm st}$ where $\phi_x^{\rm st} = (-1)^{\sum_\mu x_\mu} \phi_x$, there is also a continuous transition from the PM phase to an antiferromagnetic (AM) phase at $\kap = - \kap_{\rm PM-AM} = -0.15$. 

At non-zero $\kt$, the reduced model is still symmetric under $\kap \rightarrow -\kap - 32 \kt$, $\kt \rightarrow \kt$, and $\phi_x \rightarrow \phi_x^{\rm st}$. At small $\kt$, it is reasonable to expect the phase structure to remain similar to that at $\kt=0$ with continuous FM-PM and PM-AM phase transitions, except that $\kap_{\rm FM-PM}$ and $\kap_{\rm PM-AM}$ would now depend on the value of $\kt$. Analytic and numerical methods \cite{Bock1997fu} yield results that are consistent with this expectation. As one approaches the FM-PM transition from the FM-side, the dimensionless vacuum expectation value $|\langle \phi_x \rangle | = a |\langle \Phi_x \rangle | = v$ decreases (where $a$ and $\Phi$ are respectively lattice spacing and scalar field, both in physical units), and as a result, a radial mode (dimensionful) is developed dynamically and the unphysical {\it lgdof} are manifestly present in the continuum limit as usual scalar fields. In the reduced limit of lattice chiral gauge theories with fermions that break chiral symmetry explicitly on the lattice (e.g., Wilson fermions), the scalars couple to the fermions at such a phase transition through an effective Yukawa coupling and essentially spoil the chiral nature of the theory. The above is an undesirable outcome, and essentially leads to the failure of a large class of lattice chiral gauge theory proposals \cite{BockDeSmit91,GoltPetchSmit91}. If a lattice chiral gauge theory fails to produce chiral spectrum in the reduced limit, there is no hope in the full theory ($g\ne 0$ ) with the physical gauge fields back in the action. 

The key idea of the non-perturbative gauge fixing proposal for the abelian case is to give rise to a new universality class where the unphysical degrees of freedom ({\it lgdof})  would decouple from the physical degrees of freedom in the continuum limit. From the development so far, it appears that the large $\kt$-region is the place to look for such a possibility. For the \emph{lgdof} to decouple from the physical sector, the desired new universality class in the large $\kt$ region is to be identified with restoration of the original (target) $g$-symmetry (\ref{g-sym}). As has been found by WCPT around $g=0$ and large $\kt$, and by doing numerical simulations at weak gauge couplings \cite{Bock_etal2000}, this happens at the FM-FMD transition and the spectrum of the continuum theory, achieved by approaching the FM-FMD transition from the FM-side, contains \emph{only} free massless photons.

Following \cite{GoltSham97}, we can gain useful insight into the phase diagram in the region of small $g$ and large $\kt$ by doing a simple-minded calculation. We start from the action (\ref{S}) in the so-called vector picture, and use the property that the action has an absolute minimum at $U_{x\mu}= \exp{(iagA_\mu(x)}=1$. Near this point, the action can be expanded in powers $g$ in the constant field approximation, i.e., by neglecting derivatives of the gauge field. This leads to an expression for a classical potential density in powers of the gauge coupling $g$:
\be \label{V-cl}
V_{\rm cl} (A_\mu) = \kap \left (g^2\sum_\mu A_\mu^2 + ... \right ) + \frac{g^6}{2}\kt \left \{ \left( \sum_\mu A_\mu^2 \right ) \left ( \sum_\mu A_\mu^4 \right ) + ... \right \},
\ee
where terms with higher powers of $g^2$ are indicated by the ellipses. The classical potential density is expected to be a reasonable approximation at small $g$. However, as it turns out from numerical simulations, the classical potential density (\ref{V-cl}) produces a good qualitative picture of the new universality class in regions of the parameter space where the gauge coupling $g$ is not very small and $\kt$ is only sufficiently large, depending on the value of $g$.   

Inspection of the expression for $V_{\rm cl}$ (\ref{V-cl}) immediately leads to a critical surface defined by
\be
\kap \equiv \kap_{\rm FM-FMD}(g, \kt) = 0,
\ee
where the gauge boson (photon) is rendered massless.

Minimisation of $V_{\rm cl}$ (\ref{V-cl}) with respect to $gA_\mu$ shows that the classical potential density has two different minima at $gA_\mu = 0$ for $\kap\ge \kap_{\rm FM-FMD}$, and at
$gA_\mu = \pm \left ( \frac{|\kap - \kap_{\rm FM-FMD}|}{6\kt}\right )^{1/4}$ for $\kap < \kap_{\rm FM-FMD}$. Hence, in the quantum theory at small $g$ and large $\kt$, it is expected that tuning $\kap$ to $\kap_{\rm FM-FMD}(g,\, \kt)$ signals a new \emph{continuous} phase transition, within the {\it broken} phase, with a vector condensate as an order parameter:
\be
\langle gA_\mu \rangle &=& \pm \left ( \frac{|\kap - \kap_{\rm FM-FMD}|}{6\kt}\right )^{1/4}, \;\;\forall \mu\;\;\; {\rm for} \;\; \kap < \kap_{\rm FM-FMD} \\
\langle gA_\mu \rangle &=& 0, \;\;\forall \mu \;\;\; {\rm for} \;\; \kap\ge \kap_{\rm FM-FMD}
\ee  
The phase with the vector condensate is the novel phase and is called Ferromagnetic Directional (FMD) phase across all versions of the theory, including the theory in the reduced limit. Obviously the FMD phase breaks the rotational symmetry, and no Lorentz covariant continuum limit is obtainable from within the FMD phase. Hence, continuum limit is to be taken by approaching the continuous FM-FMD transition from the so-called Ferromagnetic (FM) phase. 

Earlier investigations done in \cite{Bock_etal2000, Bock1997fu} at weak gauge couplings are consistent with the above picture. 

It is worth mentioning here that the unfixed compact $U(1)$ lattice gauge theory, given by only the Wilson plaquette term (\ref{W}), is known to produce a phase transition at gauge coupling $g\sim 1$ between a so-called Coulomb phase containing massless free photons and a phase with non-trivial properties like having confined gauge-balls in the spectrum. The phase transition, upon precision numerical studies, was revealed as a weak first order \cite{U1late1, U1late2, U1late3}, and hence a quantum continuum limit does not strictly exist only with Wilson plaquette action. As we shall find out in Sec~\ref{Phases}, in the gauge-fixed theory under investigation with an expanded parameter space, while increasing the gauge coupling from $g<1$ to $g>1$, there is an emergence of a tricritical line at $g\sim 1$ separating a surface of continuous FM-FMD transition from a first order FM-FMD transition. The continuum limit obtained in the FM phase while approaching the continuous part of the FM-FMD transition even in the large gauge coupling ($g>1$) region would be found to consist of free massless photons only (with the \emph{lgdof} decoupled).
                 
\section{Implementation of the HMC algorithm} \label{HMC}
As indicated in the Introduction, we have written codes and tried both the MM and the HMC algorithms for generating the gauge field configurations. The MM was usually tried with 4 hits; however, various other values of hits were also tried, with very similar outcome. 

In the following, we discuss implementation of the HMC algorithm. We skip any discussion on the Wilson plaquette term of the action, because that part is standard.  
 
Writing the gauge link as $U_{x\mu}=\mathrm{exp}(i\theta_{x\mu})$, where $0<\theta_{x\mu}\le 2\pi$ is an angle (dimensionless), the HD gauge-fixing term (\ref{GS}) in the action (\ref{S}) is expressed as follows
 \begin{align}
  S_{\rm GS} = \sum_{x\mu\nu}
& \Big( \cos(\theta_{x\mu}-\theta_{x\nu}) + \cos(\theta_{x-\mu,\mu}-\theta_{x-\nu,\nu}) + 2\cos(\theta_{x\mu}+\theta_{x-\nu,\nu}) \Big) \notag \\
& - \sum_{x\mu}32\cos\theta_{x\mu} - \frac{1}{16} \sum_x\left(\sum_\mu(\sin\theta_{x\mu}+\sin\theta_{x-\mu,\mu})^2\right)^2 \notag \\ 
&+ \text{constant terms} \label{GS-long}
 \end{align}
 
The HMC algorithm, as is well known, employs a  molecular dynamics trajectory, followed by a Metropolis accept/reject step that makes the algorithm exact. The molecular dynamics trajectory is an evolution in a fictitious time (computer time) of the system through a Hamiltonian that treats the fields of the action as generalised co-ordinates and includes quadratic terms for the conjugate momenta corresponding to the fields. The Hamilton's equations of motion are discretised and the equations for momenta update involves the force terms which are the derivatives of the action with respect to the corresponding field variable.
 
Contribution to the HMC force by the HD gauge-fixing term, calculated from (\ref{GS-long}) for the field $U_{x_0\rho}$, the link field directed from the site $x_0$ towards the neighbouring site $x_0+\rho$ along the direction $\rho$, is given as
\begin{align}
-F^{\rm GS}_{x_0\rho} = & \, \frac{\partial S_{\mathrm GS}}{\partial \theta_{x_0 \rho}} \\
= &\;2 \kt \sum_\nu\Big(\sin(\theta_{x_0\nu}-\theta_{x_0\rho}) + \sin(\theta_{x_0+\rho-\nu,\nu}-\theta_{x_0\rho}) \notag\\
&- \,\sin(\theta_{x_0\rho}+\theta_{x_0-\nu,\nu}) - \sin(\theta_{x_0\rho}+\theta_{x_0+\rho,\mu})\Big) \notag\\
&+ \;32 \kt \,\sin\theta_{x_0\rho} \notag\\ 
&- \,\frac{\kt}{4}\,\cos\theta_{x_0\rho}\sum_\nu\Big( (\sin\theta_{x_0\rho}+\sin\theta_{x_0+\rho,\rho})(\sin\theta_{x_0+\rho-\nu,\nu}+\sin\theta_{x_0+\rho,\nu})^2 \notag\\
&+ \,(\sin\theta_{x_0-\rho,\rho}+\sin\theta_{x_0,\rho})(\sin\theta_{x_0-\nu,\nu}+\sin\theta_{x_0\nu})^2\Big)\notag
\end{align}
 
 A similar contribution to the force from the dimension-2 mass counter-term is easily found to be
\be
-F^{\rm ct}_{x_0\rho} = 2\kap \sin\theta_{x\mu}.
\ee
  
\subsection{Some details of our numerical simulations}
Numerical simulations were carried out to generate statistically independent gauge field configurations at gauge couplings $g = 0.6, \, 0.8, \, 1.0, \, 1.1, \, 1.2, \, 1.3, \,1.4, 1.5, \, 1.6$ and 1.8, at different lattice volumes such as $8^4$, $10^4$, $12^4$, $16^4$, $20^4$, $24^4$, $8^3 24$ and $10^3 24$. In this paper, we present results at gauge coupling $g = 1.0, \, 1.1, \, 1.2, \, 1.3$ and $1.5$ and at lattice volumes $8^4$, $10^4$, $12^4$, $16^4$ and $8^3 24$, because these had the most statistics and the set was deemed enough to establish our conclusions. However, data at all other gauge couplings and volumes, especially bigger volumes, while not having the same refinement level as the ones presented in this paper because of lower statistics, help in some way for double-checking the conclusions made in this paper. At every gauge coupling and lattice volume, the ($\kt, \, \kap$) parameter space was scanned by independent Monte Carlo runs in both directions. To locate phase transitions precisely, these runs around the phase transitions were performed in fine steps of $\Delta \kap =0.001$ and $\Delta \kt=0.005$. Typically each run for the data presented here at a given $g, \, \kt, \, \kap$ had at least 20,000 equilibrated trajectories/sweeps at each lattice volume. Integrated autocorrelation times were calculated for each measured quantity by the well-known window method. Error bars have been calculated taking the autocorrelations into account. Error bars have been omitted wherever they are smaller than the symbols used.  

Vacuum expectation values of quantities that were measured on equilibrated gauge field configurations on $L^4$ (or $L^3 T$, $L\ne T$ for propagators) lattices, are the average plaquette 
\be
E_{\rm P} = \frac{1}{6L^4} \left \langle \sum_{x,\mu<\nu} {\rm Re}\, U_{{\rm P}\mu\nu}(x) \right \rangle, 
\ee
the gauge field mass term 
\be
E_\kappa = \frac{1}{4L^4} \left \langle \sum_{x,\mu} {\rm Re}\,U_{x\mu} \right \rangle, 
\ee
and the lattice version of the vector condensate $\langle A_\mu \rangle$ 
\be \label{V}
V = \left \langle \sqrt{\frac{1}{4}\sum_\mu \left(\frac{1}{L^4} \sum_x {\mathrm{Im}}  \, U_{x\mu}  \right )^2} \, \right \rangle .
\ee
The vector condensate $V$ is the order parameter for the FM-FMD transition. It is zero for all other phases except FMD. For first order FM-FMD transition, the quantity $E_\kap$ goes through a finite jump when plotted against $\kap$. However, for continuous FM-FMD transition, the finite jump vanishes and $E_\kap$ is continuous across the transition. 

To locate and determine the order of the phase transitions involving FMD, e.g., FM-FMD, AM-FMD and PM-FMD, the observables $V$ and $E_\kap$ are very useful. To determine the location on a finite lattice, a suitable criterion has to be set. For all continuous phase transitions in our investigation, the location with the highest fluctuations in the data is taken as the approximate position of the phase transition. For all first order transitions, the standard histogram method, as used by us previously (e.g., look at Fig.~3 in \cite{DeSarkar2016}), was employed. 
The location of the FM-PM and PM-AM phases are harder to determine, staying within the observables of the theory in the vector picture. However, we find that given the definition of $V$ strictly as a positive quantity as in (\ref{V}), the increased fluctuations of the fields around these continuous transitions are captured quite precisely around the phase transitions by the quantity $V$ even though neither of these phases FM, PM and AM have a vector condensate.
Out of all the phase transitions to be presented later in this paper, the one of prime importance to us is the FM-FMD, and the location and nature of this transition including the tricritical points naturally attracted most of our attention.

In addition, vector propagators
\be \label{gauge-prop}
G_{\mu\nu}(p) = \frac{1}{g^2 L^3T} \left \langle \sum_{x,\,y} {\mathrm{Im}}\,U_{x\mu} \,{\mathrm{Im}}\,U_{y\nu} \,{\mathrm{exp}} [ip(x-y)] \, \right \rangle
\ee
and effective scalar propagators 
\be \label{scalar-prop}
H_{\mu\nu}(p) = \frac{1}{L^3T} \left \langle \sum_{x,\,y} {\mathrm{Re}}\,U_{x\mu} \,{\mathrm{Re}}\,U_{y\nu} \,{\mathrm{exp}} [ip(x-y)] \, \right \rangle
\ee
were computed in momentum space as functions of the allowed momenta $p$ on periodic lattices of volume $L^3T$ with $L$ and $T$ respectively as spatial and temporal extensions. The operator ${\mathrm{Re}} \, U_{x\mu}$ carries quantum numbers of a scalar field and the expression given in Eq.~(\ref{scalar-prop}) was used in the past in gauge-Higgs systems to compute Higgs mass \cite{Evertz_etal1987}.

The FM-FMD transition was also probed with vectorially coupled quenched Kogut-Susskind fermions having $U(1)$ charge, by measuring the chiral condensate
\be
\langle  \overline{\chi} \chi  \rangle_{m_0} = \frac{1}{L^4} \sum_x \left \langle M^{-1}_{xx} \right \rangle 
\ee
as a function of a vanishing bare fermion mass $m_0$. The fermion matrix $M$ is given by,
\be
M_{xy} = \frac{1}{2} \, \sum_{\mu=1}^4 \eta_\mu(x) \left ( \delta_{y, x+\mu} U_{x\mu} - \delta_{y, x-\mu} U^\dagger_{x-\mu,\mu} \right ) + m_0 \, \delta_{x,y},
\ee
with $\eta_\mu(x)\equiv (-1)^{x_1+...+x_{\mu-1}}$ and $\eta_1(x)\equiv 1$.
Noisy estimator method \cite{Bitar_etal1989} with four noise vectors was used to compute the chiral condensate. Anti-periodic boundary conditions were used for the quenched fermions and the fermion matrix $M$ was iteratively inverted using the standard Conjugate Gradient (CG) inverter. A more modern inverter, viz., BiCGStab was also tried, with no gains for the number of iterations needed for convergence; however, it had substantial computational overhead compared to CG and hence not used.

\section{Local versus Global Algorithm} \label{Algorithms}
\begin{figure}
\begin{center}
\includegraphics[width=0.48\linewidth]{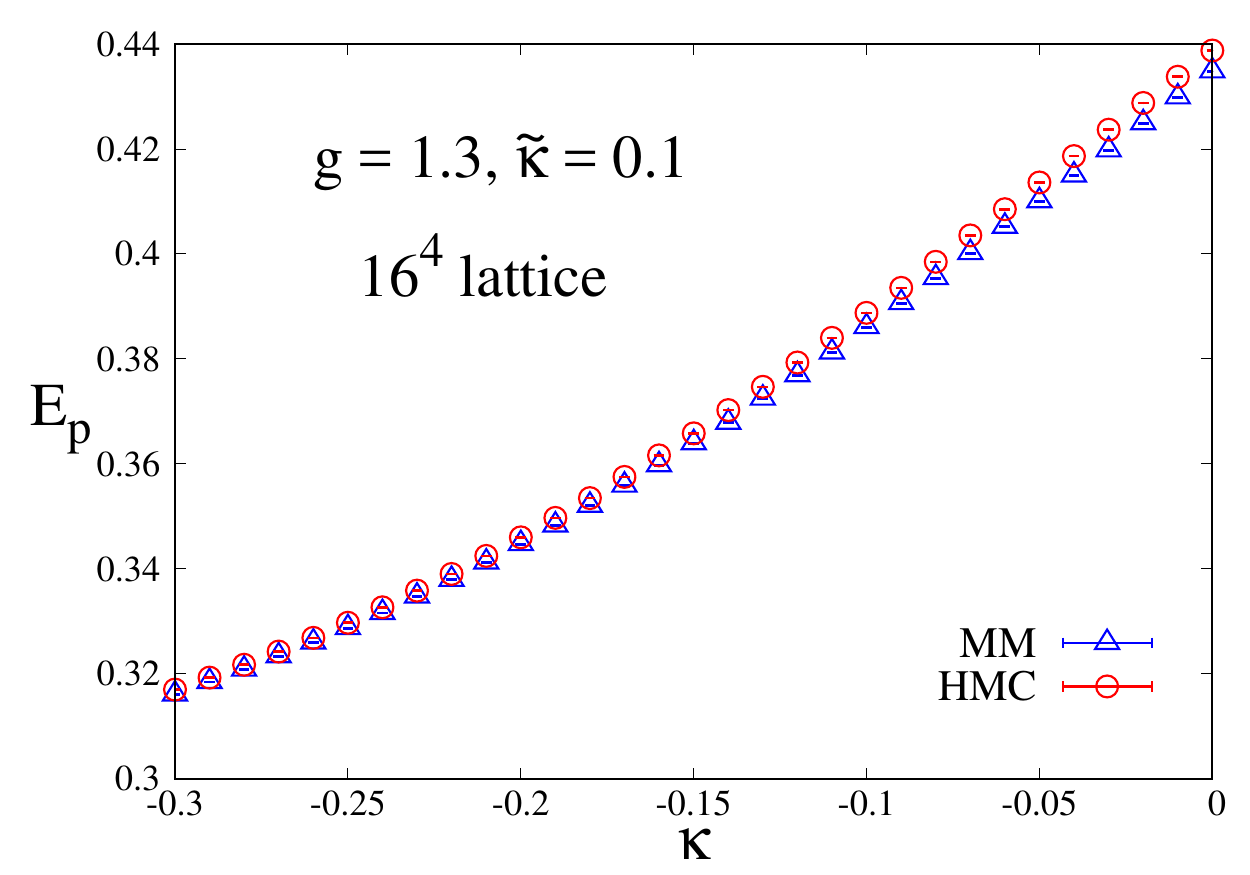}  \quad
\includegraphics[width=0.48\linewidth]{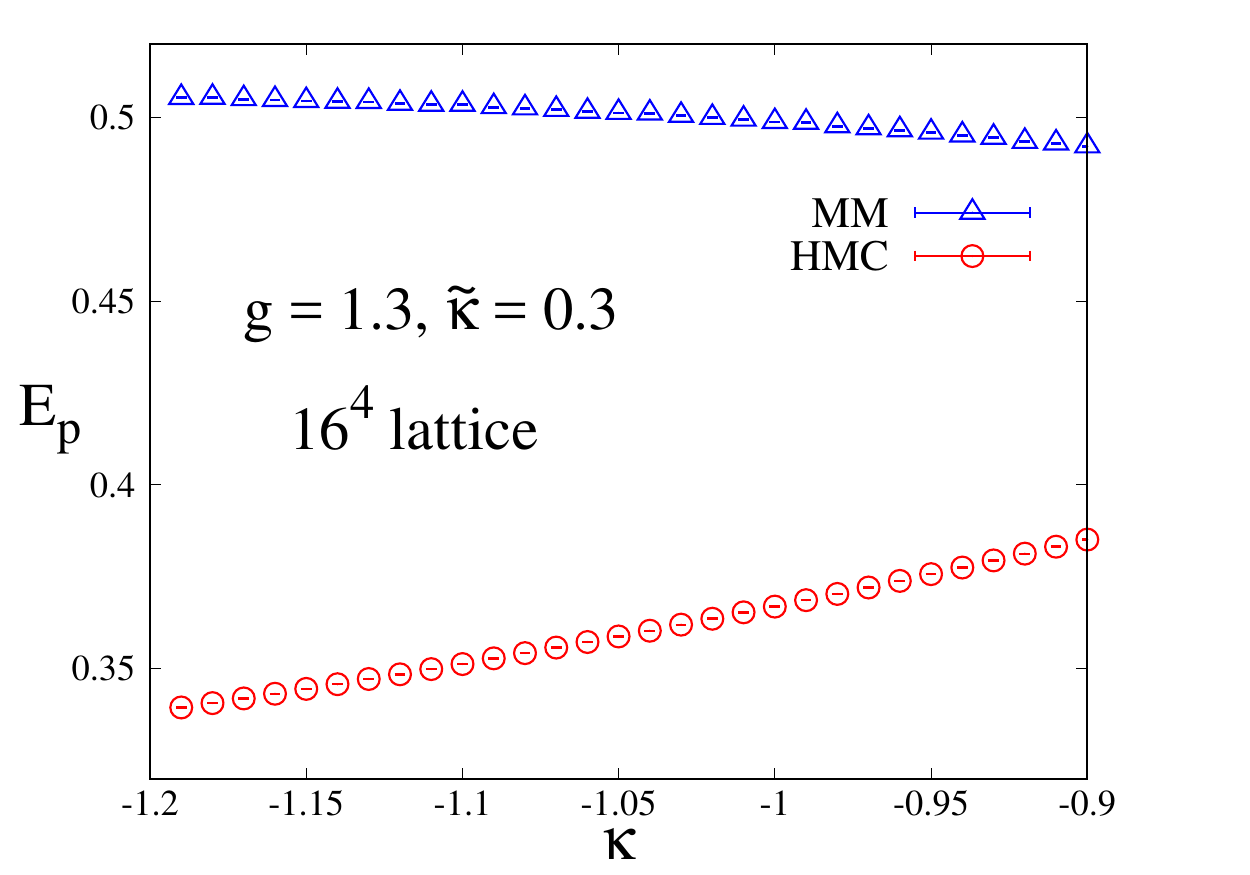}
\end{center}
\caption{Comparison of average plaquette value $E_p$ at  $g=1.3$ on $16^4$ lattice at two values of $\kt$ (= 0.1 and 0.3) in the two figures for a variety of values of $\kap$ around the FM-PM transition, obtained with MM and HMC algorithms.}
\label{MMvsHMC1}
\end{figure}
\begin{figure}
\begin{center}
\includegraphics[width=0.48\linewidth]{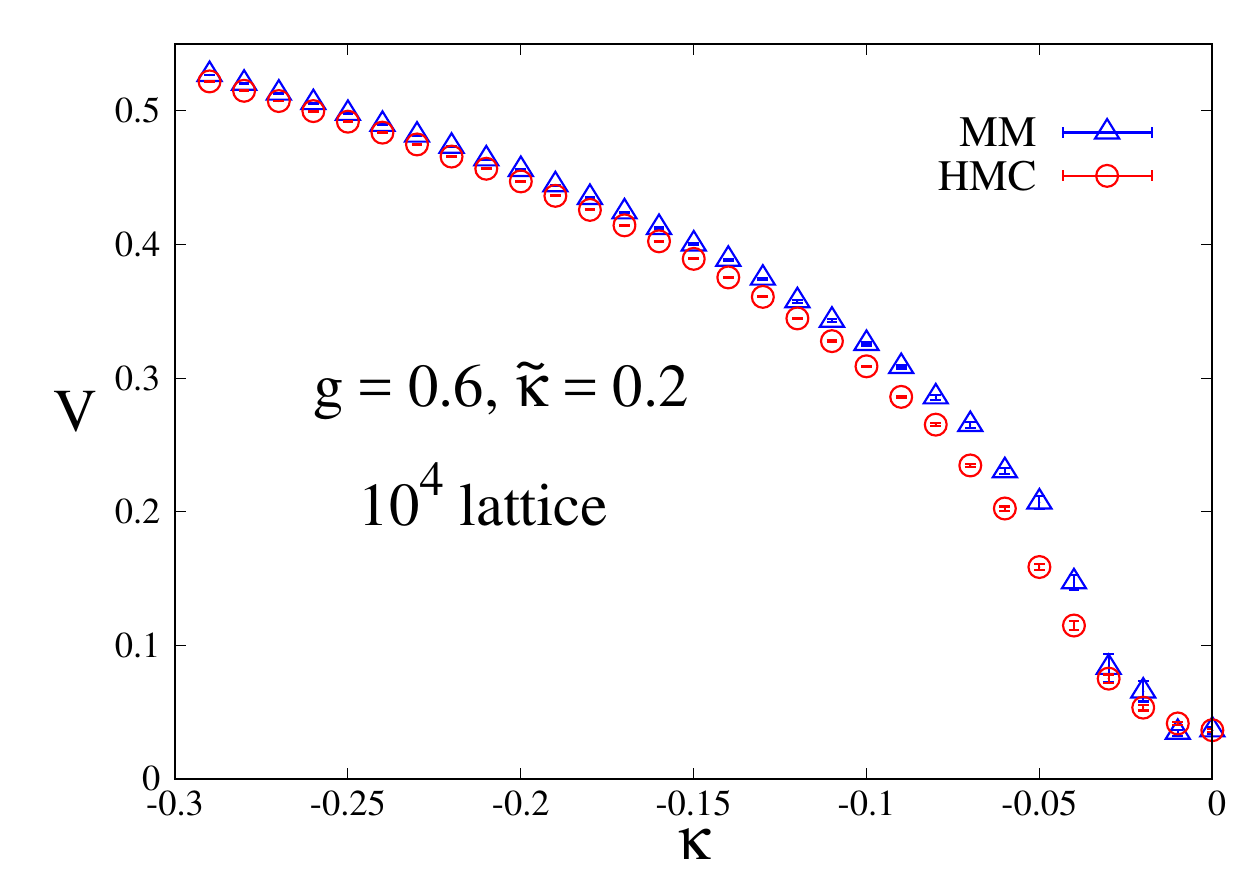}  \quad
\includegraphics[width=0.48\linewidth]{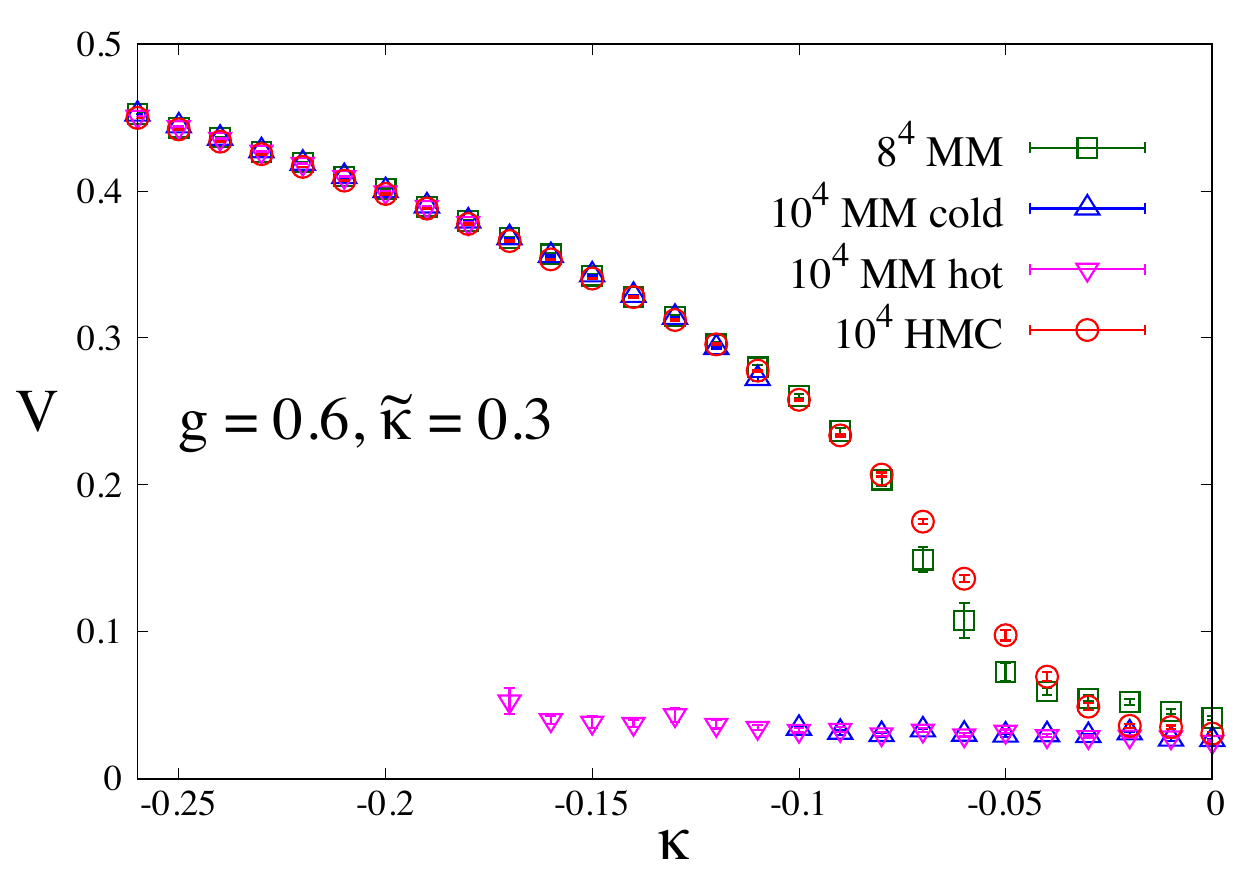}
\end{center}
\caption{Comparison of the vector condensate $V$ at  $g=0.6$ on $10^4$ lattice at two values of $\kt$ (= 0.2 and 0.3) in the two figures for a variety of values of $\kap$ around the FMD-FM transition, obtained with MM and HMC algorithms. The right figure also has values with MM algorithm on $8^4$ lattices.}
\label{MMvsHMC2}
\end{figure}

In this section, we present results that show that a local algorithm like MM appears to be unreliable when the coefficient of the HD gauge-fixing term $\kt$ is relatively large. Since the continuous part  of the phase transition of interest (FM-FMD) takes place at a larger value of $\kt$ when the gauge coupling is stronger, the problem is more apparent in the current work since it primarily deals with the fate of the theory at strong gauge couplings.  We also show that the results are unstable as the volume increases. 

In contrast, the HMC algorithm, a so-called global algorithm, generally appears to be more reliable and consistent, and undoubtedly a better algorithm for this theory with a HD term. First, it generally agrees with results in \cite{Bock_etal2000} obtained in the weak gauge coupling region for rather small volumes, mostly $4^4$ and $6^4$ and some data for $8^4$. At larger $\kt$, both for weak and strong gauge couplings, and for larger volumes, $10^4$ and above, HMC gives stable results, as will be illustrated in the following. 

\begin{figure}
\begin{center}
\includegraphics[width=0.66\linewidth]{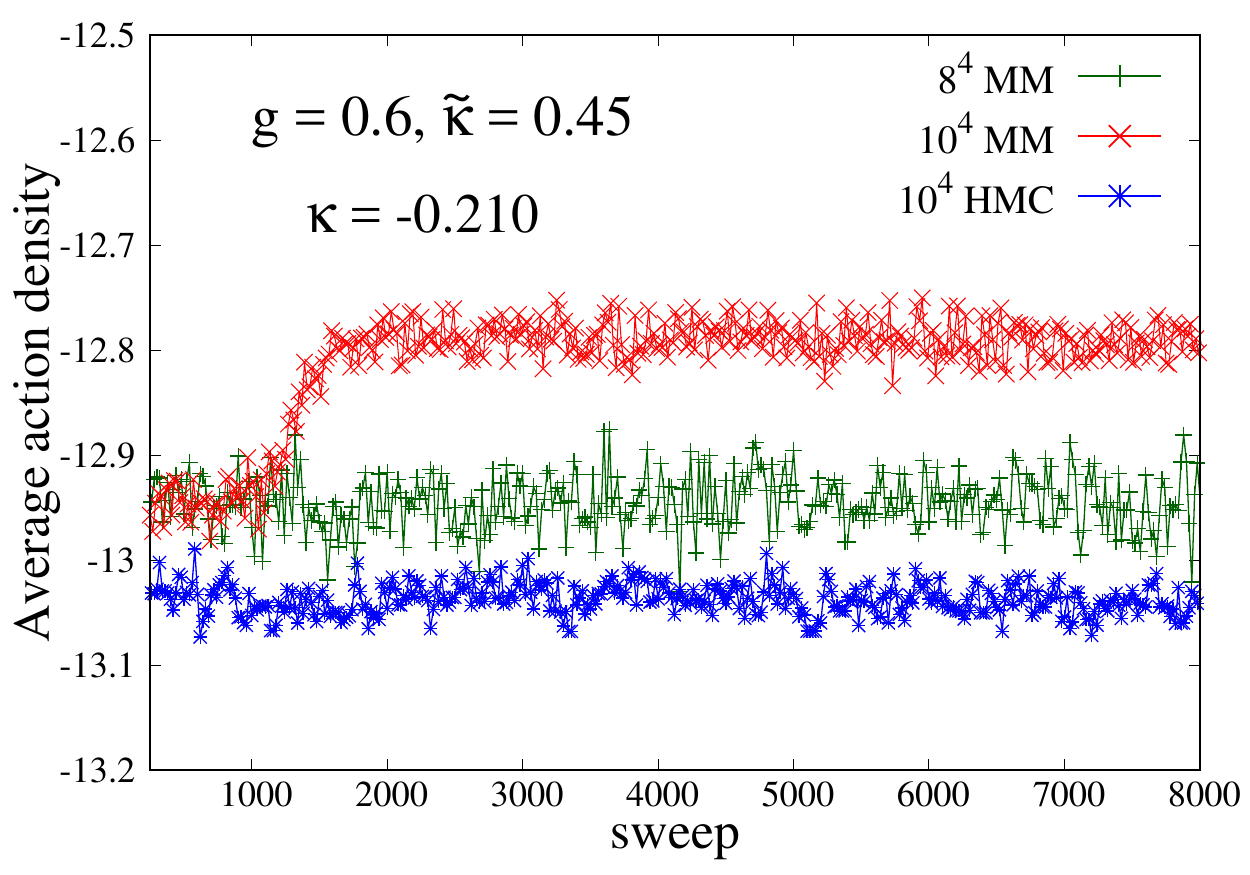} 
\end{center}
\caption{Average action density versus number of sweeps/trajectories just inside the FMD phase at a weak gauge coupling $g = 0.6$ with MM and HMC algorithms.}
\label{MMvsHMC3}
\end{figure}

In Fig.~\ref{MMvsHMC1}, we have compared average plaquette value $E_p$ at  a \emph{strong} gauge coupling $g=1.3$ on $16^4$ lattice at two values of $\kt$ (= 0.1 on the left and 0.3 on the right) for a variety of values of $\kap$ around the FM-PM transition, obtained with MM and HMC algorithms. The observable $E_p$ does not show any noticeable signal for the transition. However, for $\kt = 0.1$, the two algorithms produce nearly identical results for all values of $\kap$ presented, while for the larger $\kt = 0.3$, results given by the two algorithms have no agreement anywhere.

Fig.~\ref{MMvsHMC2} shows comparison of the observable $V$, an estimate of the vector condensate $\langle A_\mu(x) \rangle $ on the lattice, obtained with MM and HMC algorithms, at a \emph{weak} gauge coupling $g=0.6$ and two values of the coefficient $\kt$ of the HD gauge-fixing term. On $10^4$ lattices, we see general agreement of the two algorithms at smaller $\kt$ in the left figure. However, for the same $10^4$ lattices, the algorithms clearly give different results (indicating also a dependence on the initial configuration for the MM) at a slightly larger value of $\kt=0.3$ (in the right figure), suggesting a first-order-like discrete jump in the quantity $V$ at a shifted location of the parameter $\kap$, while the MM data on $8^4$ lattices generally agree with the $10^4$ HMC data signalling a smooth transition.    

Average action density (value of action divided by volume) achieved after apparent equilibration is plotted in Fig.~\ref{MMvsHMC3} against number of sweeps/trajectories at a weak gauge coupling $g=0.6$ with both MM and HMC algorithms at a point in the coupling parameter space just inside the FMD phase. While the system settles at the lowest average action density with the HMC algorithm, the MM algorithm clearly produces unreliable results with the average action density above that obtained with HMC and showing instability with change of volume.  

In fine, our findings are that, the MM algorithm struggles to generate reliable gauge field configurations with the HD action. It produces correct results only for sufficiently small coefficient $\kt$ of the HD gauge-fixing term and on small lattices. The situation gets particularly worse on larger lattices, and at strong gauge couplings where one needs a large $\kt$ for a continuous transition. 

In contrast, the HMC algorithm agrees with the MM results at small values of $\kt$ and small lattice volumes. In addition, the results with the HMC appear more consistent and stable with change of lattice volume and parameters of the algorithm. However, at times even the HMC can struggle with this HD action to move out of a local metastability because the changes of the fields and the momenta along a molecular dynamics trajectory are tiny with every differential `time'-step. We have found it beneficial to use initially the MM algorithm for any configuration-generating run at a given point in the parameter space, and then feed the final configuration of the MM-run as the initial configuration of the HMC run, to make use of the best of both algorithms. This is because, even though the MM algorithm is a local algorithm, meaning the Metropolis accept/reject step is performed after each change of the gauge field at a given link, the changes of the values of the gauge fields are finite, unlike the differential changes during a trajectory of the HMC algorithm.  

\section{Study of the Phase Diagram at strong gauge couplings} \label{Phases}
\begin{figure}
\begin{center}
\includegraphics[width=0.66\linewidth]{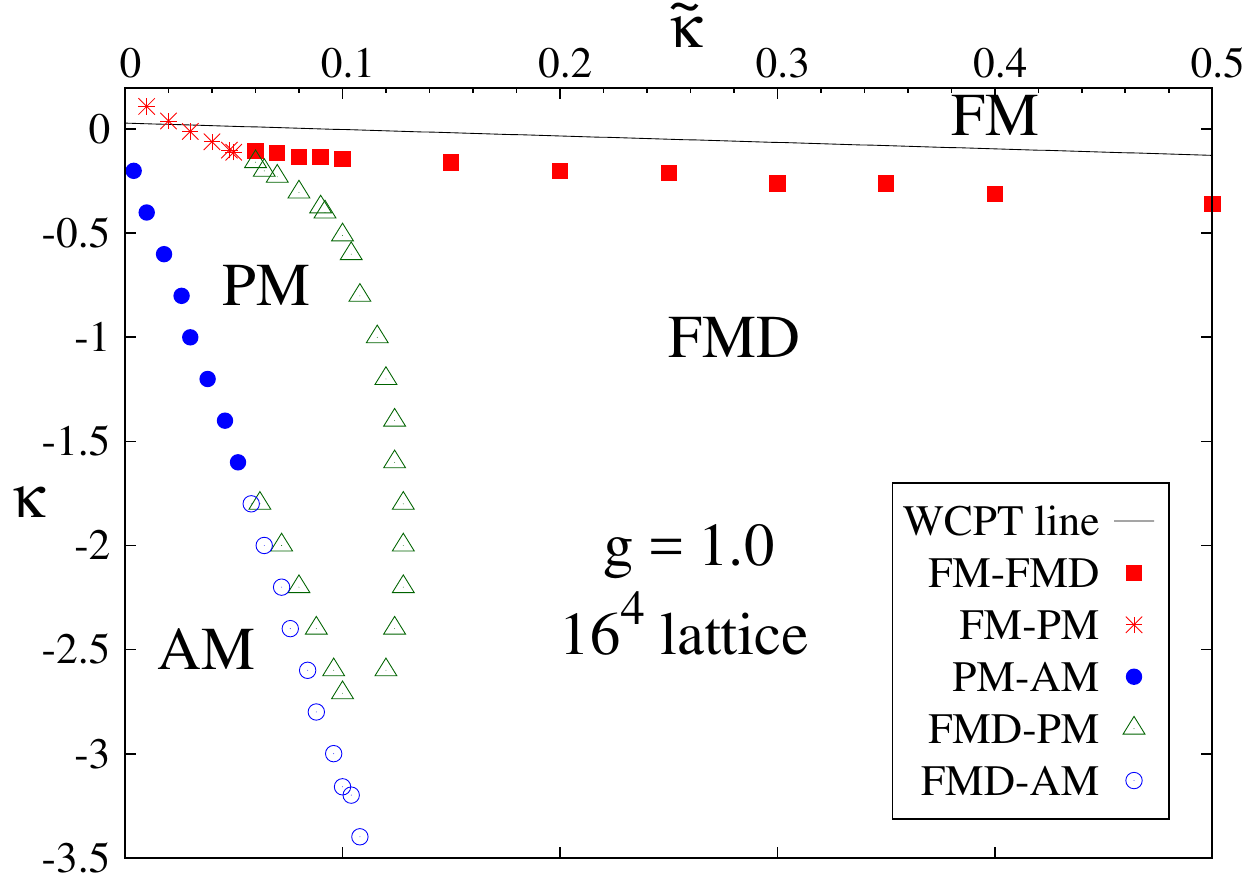} 
\end{center}
\caption{Phase diagram in the $(\kt,\kap)$ plane at gauge coupling $g=1.0$ on $16^4$ lattice.}
\label{weakphasedia}
\end{figure}
The unfixed compact $U(1)$ lattice gauge theory ($\kt=0, \; \kap=0$), as formulated by the plaquette action of Wilson and given in (\ref{W}), has been studied extensively in the past. At $g\sim 1$, the theory shows a phase transition from a so-called Coulomb phase (at weak gauge couplings, $g<1$) having massless free photons, to a phase having non-trivial properties like confinement and existence of gauge-balls etc. at strong gauge couplings ($g>1$). Through careful Monte Carlo simulations, the order of the transition was determined to be weakly first order. Hence a quantum continuum limit was not possible at this phase transition.

It is mentioned in Sec.~\ref{Intro} and \ref{Review} that the phase diagram of the theory (\ref{S}) under investigation was studied for weak gauge couplings reasonably extensively in the past. From the point of view of quantum field theory, there are two continuous transitions of interest - the FM-PM transition and the FM-FMD transition. These transitions are illustrated in the phase diagram presented in Fig.~\ref{weakphasedia} obtained in our numerical simulation at gauge coupling $g=1$, approximately the largest gauge coupling exhibiting all the features of the phase diagram at weak gauge couplings ($g < 1$). As discussed in Sec.~\ref{Review}, a gauge-scalar (popularly known as gauge-Higgs) theory is expected to emerge in the continuum limit near the FM-PM transition at small values of $\kt$. However, at larger values of $\kt$, we find that, approaching the FM-FMD transition from the FM side makes the scalar fields (\emph{lgdof}) decouple as gauge symmetry is recovered at that transition with emergence of massless free photons. The FMD phase is marked by a vector condensate, and hence approaching the FM-FMD transition from the FMD side cannot produce a Lorentz covariant theory.

In Fig.~\ref{weakphasedia} and all phase diagrams to follow, all data points represented by solid (filled) symbols signify a continuous phase transition, while all data points represented by empty (unfilled) symbols signify a first order transition. Accordingly, one would find that FM-PM and PM-AM phase transitions are continuous and PM-FMD phase transition is first order, for all gauge couplings investigated in this paper.

\begin{figure}
 \begin{center}
  \includegraphics[width=0.48\linewidth]{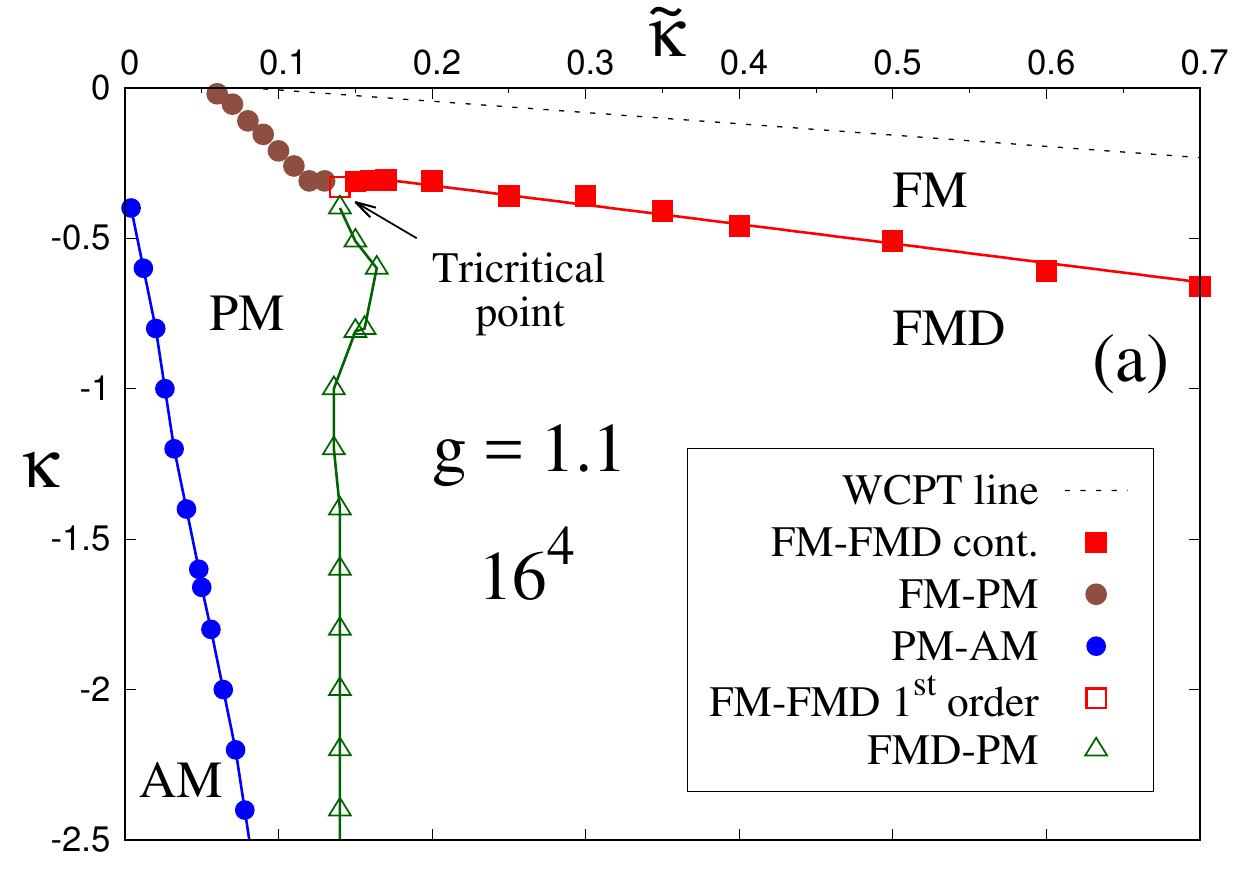} \quad
  \includegraphics[width=0.48\linewidth]{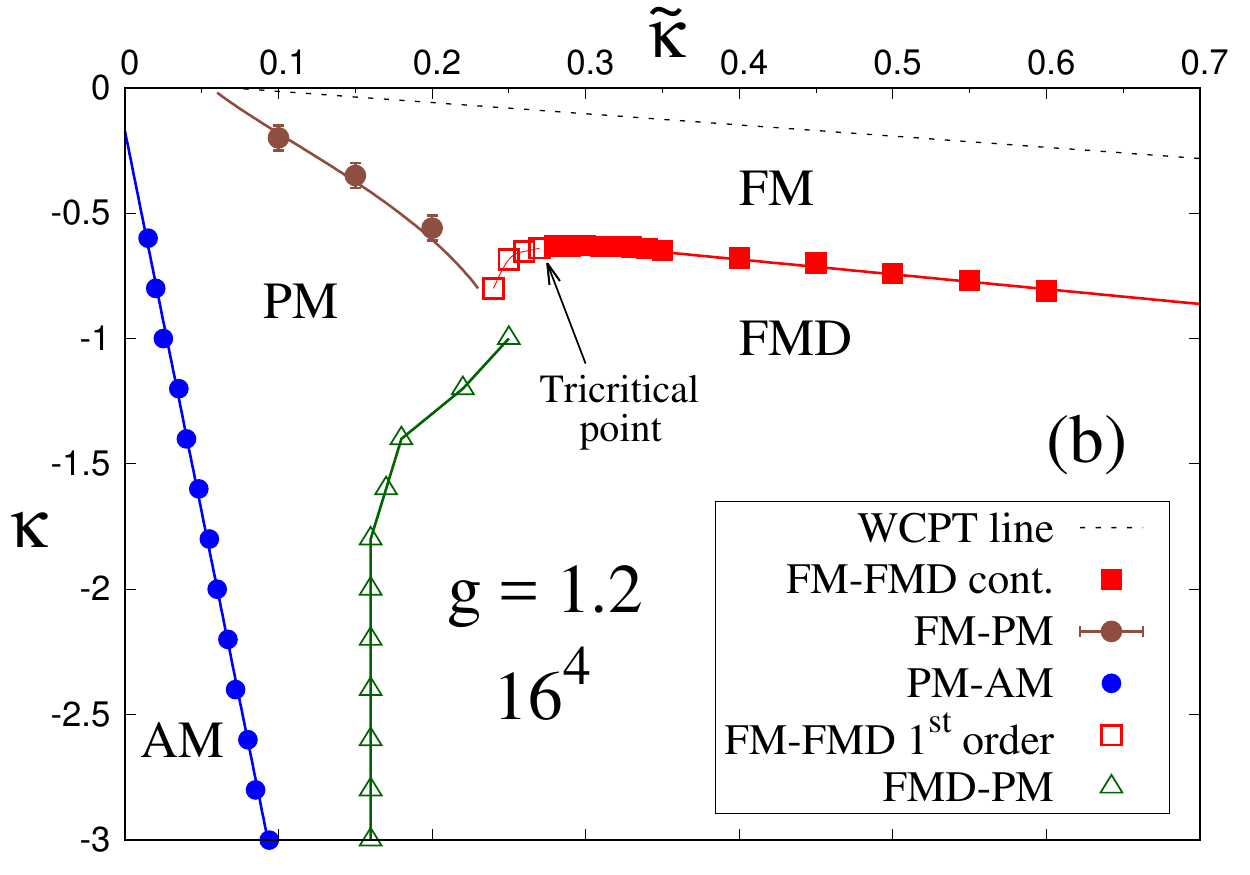} \\
  \includegraphics[width=0.48\linewidth]{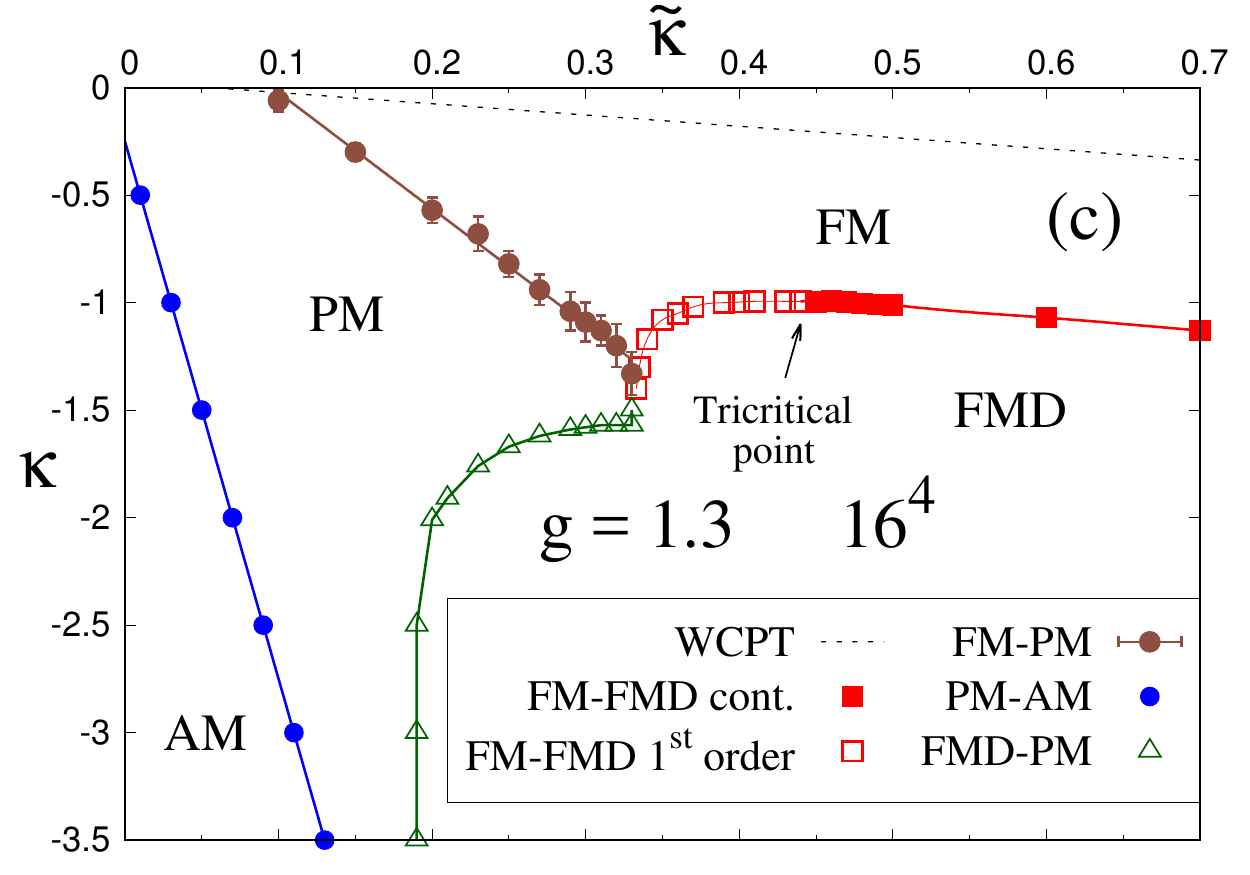} \quad
  \includegraphics[width=0.48\linewidth]{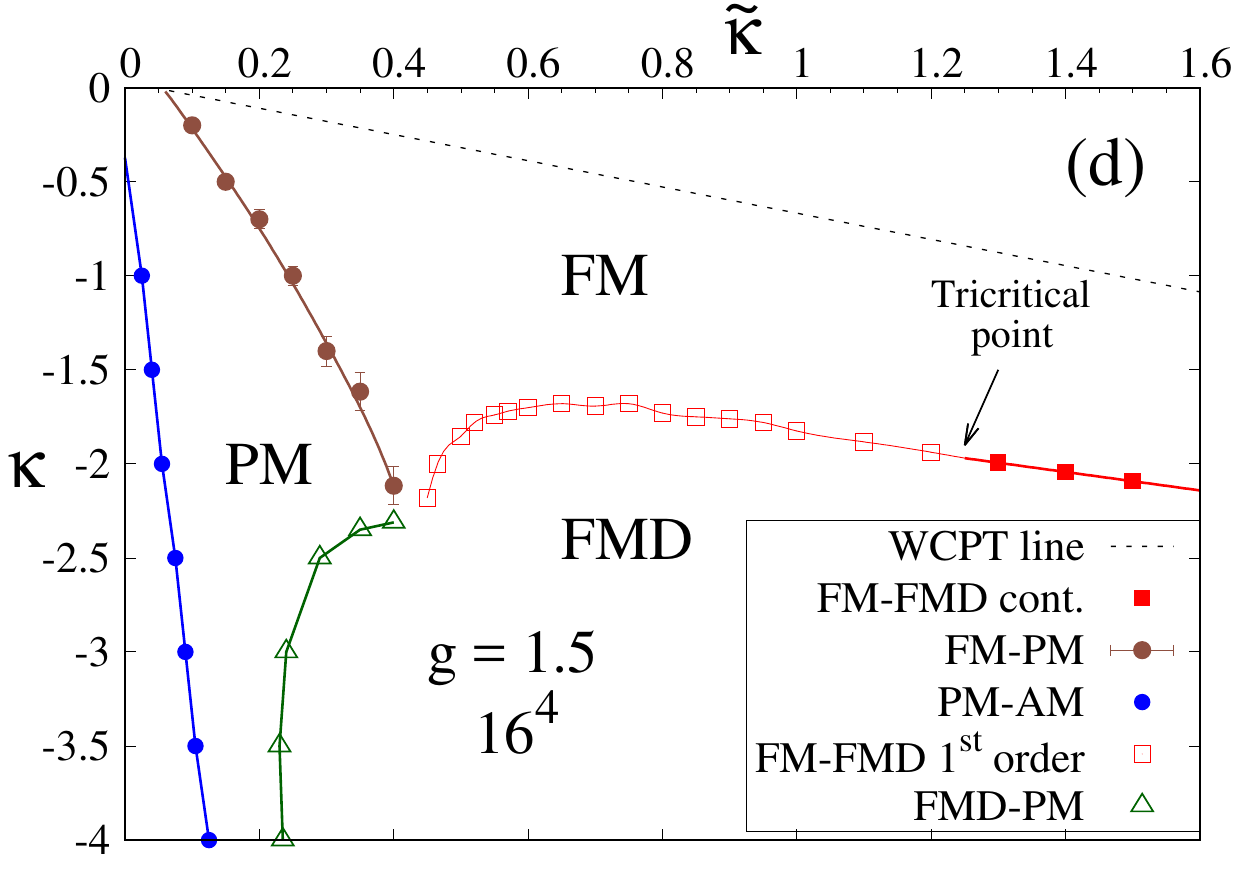}
  \end{center}
  \caption{Phase diagram in the $(\kt,\kap)$ plane at gauge coupling $g=1.1, \; 1.2, \; 1.3$ and 1.5 on $16^4$ lattices. The dotted line shows the FM-FMD continuous transition calculated from WCPT around $g=0$}
  \label{PhaseDiag}
 \end{figure}

However, for $g=1.1$ and greater values of the gauge coupling $g$, the FM-FMD phase transition develops a first order part for smaller values of $\kt$, as seen in Figs.~\ref{PhaseDiag}(a), (b), (c) and (d). At $g=1.1$ (Fig.~\ref{PhaseDiag}(a)), in our simulations on $16^4$ lattice, the FM-FMD phase transition first shows a little glimpse of its first order part for small values of $\kt$ and then quickly turns itself into a continuous transition at a tricritical point at $(\kt, \,\kap) \sim (0.14, -0.33)$ and remains continuous for larger values of $\kt$. As the gauge coupling $g$ is increased to $g=1.2$ (Fig.~\ref{PhaseDiag}(b)), $g=1.3$ (Fig.~\ref{PhaseDiag}(c)) and $g=1.5$ (Fig.~\ref{PhaseDiag}(d)), the location of the triciritical point in the ($\kt$, $\kap$)-plane shifts to larger $\kt$ and more negative $\kap$. In other words, the first order part of the FM-FMD phase transition extends quite rapidly with increase of the gauge coupling. However, it appears from our numerical simulations (which includes gauge couplings $g>1.5$, corresponding data not shown here) that, given a large gauge coupling there is always a sufficiently large $\kt$ beyond which the FM-FMD transition is continuous. In addition, the FM-FMD transition overall shifts to larger negative $\kap$ values at stronger gauge couplings. 

The line joining the discrete data points at the phase transitions in the four plots of Fig.~\ref{PhaseDiag} are simple-minded interpolations with a purpose to guide the eye. The exact location of the meeting point of the three phases FM, FMD and PM are only roughly determined for a couple of the plots. However, such points where the continuous FM-PM transition ends at the joining of two (PM-FMD and FM-FMD) first order lines (for $g>1$) have been described as critical end points in \cite{Somen}.  

At weak gauge couplings studied for example in \cite{Bock_etal2000}, the order of the continuous FM-FMD transition was expected to be second order from analytic considerations. In our case at strong gauge couplings, we conclude from numerical evidence of susceptibility peaks of the vector condensate at different volumes that the continuous part of the FM-FMD transition is also second order.  

\begin{figure} \label{3dPD}
\begin{center}
\includegraphics[width=0.75\linewidth]{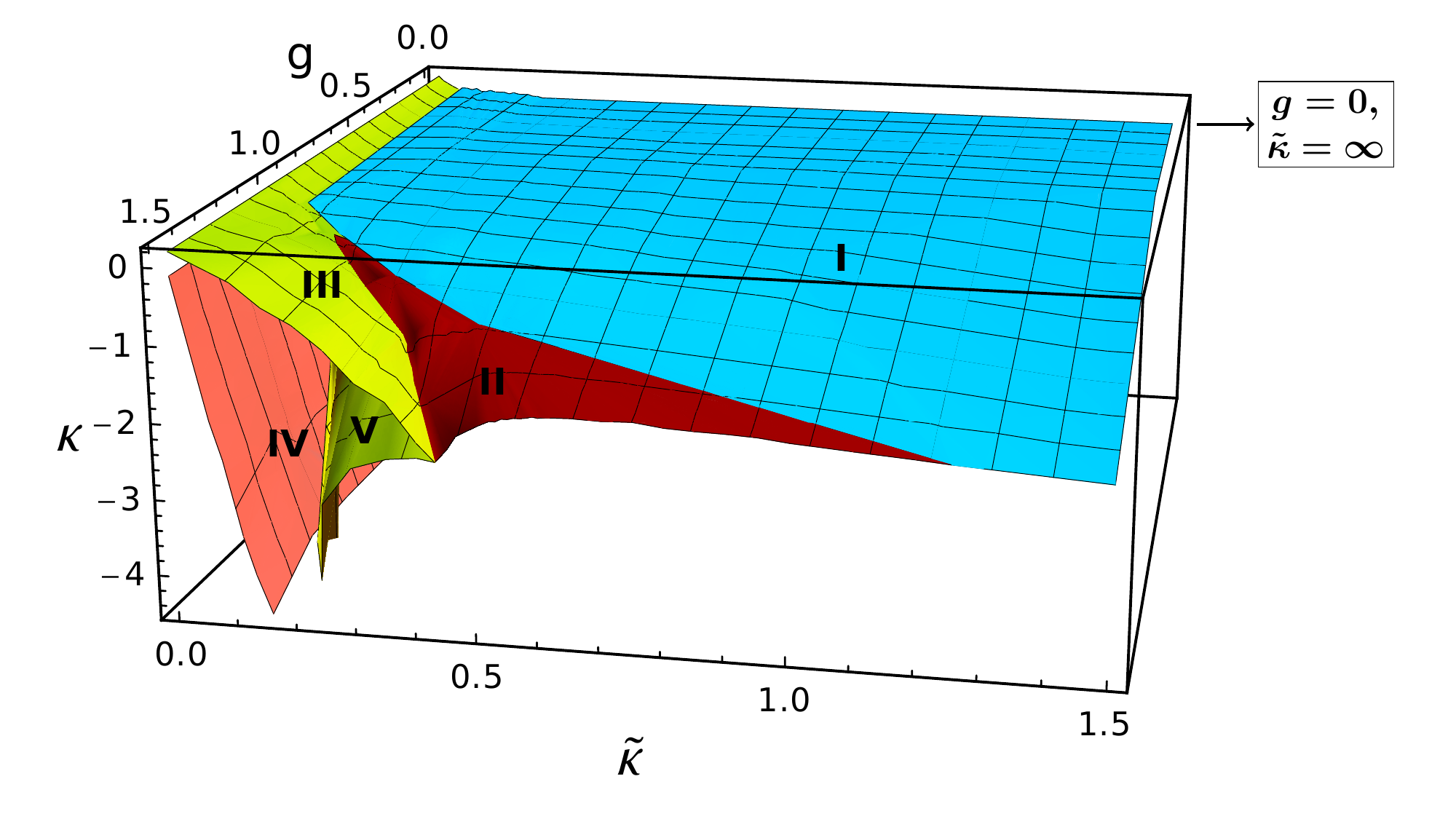}
\end{center}
\caption{Schematic phase diagram in the 3-dimensional $(g,\, \kt,\,\kap)$ parameter space at gauge coupling based on available data on $16^4$ lattice. Surfaces of different phase transitions are labelled by: I: FM-FMD (continuous), II: FM-FMD (first order), III: FM-PM (continuous), IV: PM-AM (continuous), V: PM-FMD (first order). A tricritical line separating the first order and the continuous FM-FMD transitions emerges at $g>1$ and continues to move towards larger $\kt$ at stronger gauge coupling $g$. The arrow at the top right points to the WCPT corner ($g=0, \, \kt \rightarrow \infty$).}
\label{3dpd}
\end{figure}

At $g=1.1$, where in our numerical simulations on $16^4$ lattices, the first order part of the FM-FMD transition makes an appearance for the first time, the discrete jump in the quantity $E_\kap$ across the transition is small, accordingly we conclude that the order of the transition is a weak first order. However, as the gauge coupling increases, the discrete jump in $E_\kap$ becomes quite pronounced, making the initial part (for smaller $\kt$) of the FM-FMD transition strongly first order, which then becomes progressively weaker with increase of $\kt$, until the transition line reaches the tricritical point, beyond which the transition is of course continuous.

The dotted straight line in each of the figures of Fig.~\ref{PhaseDiag} is obtained in bare WCPT near $g=0$ by demanding recovery of gauge symmetry and is representative of the FM-FMD transition in $(\kt, \,\kap)$-plane for a given gauge coupling $g$ \cite{Bock_etal2000}. The dotted lines are always nearly parallel to the continuous parts of the FM-FMD transition for all gauge couplings in Fig.~\ref{PhaseDiag}. However, the actual transitions run always lower in the $(\kt, \,\kap)$-plane, and their distance from the WCPT lines increase with increasing gauge coupling.      

In the next Section, at strong bare gauge couplings, we shall explore the physics, achievable by approaching the continuous part of the FM-FMD phase transition from the FM phase, by computing the gauge field propagator, an effective scalar field propagator and chiral condensates. However, while the bare WCPT done around the point $g=0$ and $\kt=\infty$ has limited range of applicability, there exists no phase transitions between the WCPT corner of the 3-dimensional coupling parameter space (viz., $g=0$ and $\kt=\infty$) and any point on the continuous part of the FM-FMD transition at a strong gauge coupling and a large enough $\kt$. The schematic phase diagram in the 3-dimensional parameter space ($g,\, \kt,\, \kap$) is displayed in Fig.~\ref{3dpd}. Kindly note that $\kap=0$ surface is located slightly below the top surface of the 3-dimensional box presented in the figure. The diagram is drawn based on available data on phase transitions and interpolations and extrapolations. The continuity of the entire FM-FMD transition surface (bounded by the tricritical line starting at $g=1.1$) up to the WCPT corner is clearly evident when we look at the 3-dimensional phase diagram. Hence it is natural to expect that this whole region falls under the same universality class and the continuum physics obtainable should be no different from that near the weak gauge coupling region.       
  
\section{Vector and Scalar propagators, Chiral Condensates and  the Plaquette} \label{Measurements}
\begin{figure}
\begin{center}
\includegraphics[width=0.66\linewidth]{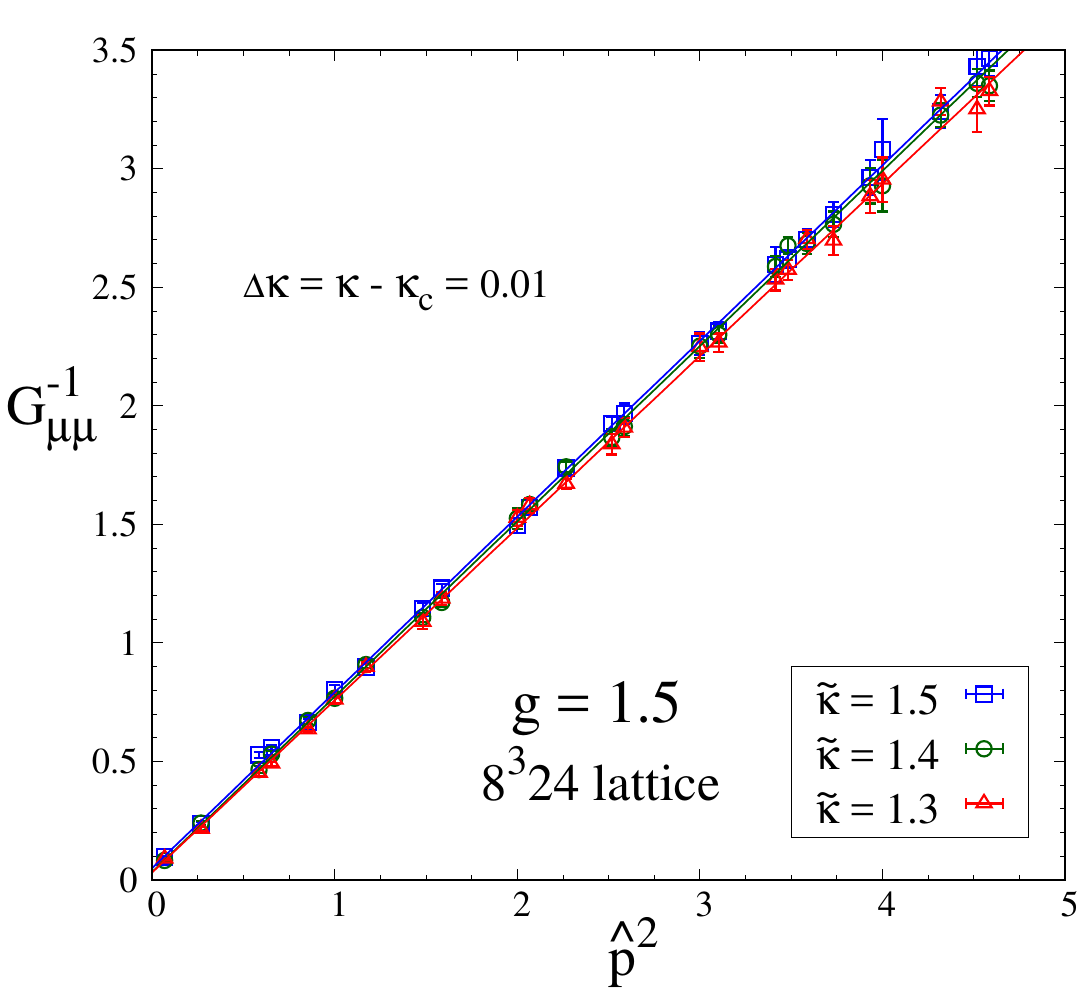}
\end{center}
\caption{The inverse of the gauge field propagator versus $\hat{p}^2$ at $g=1.5$ near the continuous FM-FMD transition on $8^324$ lattices indicating emergence of free massless photons.}
\label{gprop}
\end{figure}
The inverse of the gauge field propagator, as given by Eq.~\ref{gauge-prop}, for $\mu=\nu$, is plotted in Fig.~\ref{gprop} against the lattice momentum $\hat{p}^2 = 2 \sum_\mu \sin^2 p_\mu/2$ at $g=1.5$ and $\kt=1.3,\, 1.4, \, 1.5$ on $8^3 24$ lattices. The $\kappa$ values are chosen to stay very close to the FM-FMD transition (continuous for the above $\kt$ values). The linear behaviour of the fits passing nearly through the origin clearly indicates a vanishing photon mass at the transition. At each $\kt$ value, although not shown in the plot, the photon mass scales with decreasing $\kap$ approaching the transition from the FM side. As $\kt$ increases from 1.3 to 1.5, there is a small but monotonic increase of the slope of the fit. The corresponding figure at $g=1.3$ in \cite{DeSarkar2016} also has the same trend with increasing $\kt$. Consistency of this trend in these two figures and in data at other gauge couplings (not shown here) suggests that at larger $\kt$ the slope is likely to approach unity, in tune with theoretical expectations, rendering the photons perfectly free.  

\begin{figure}
 \begin{center}
  \includegraphics[width=0.66\linewidth]{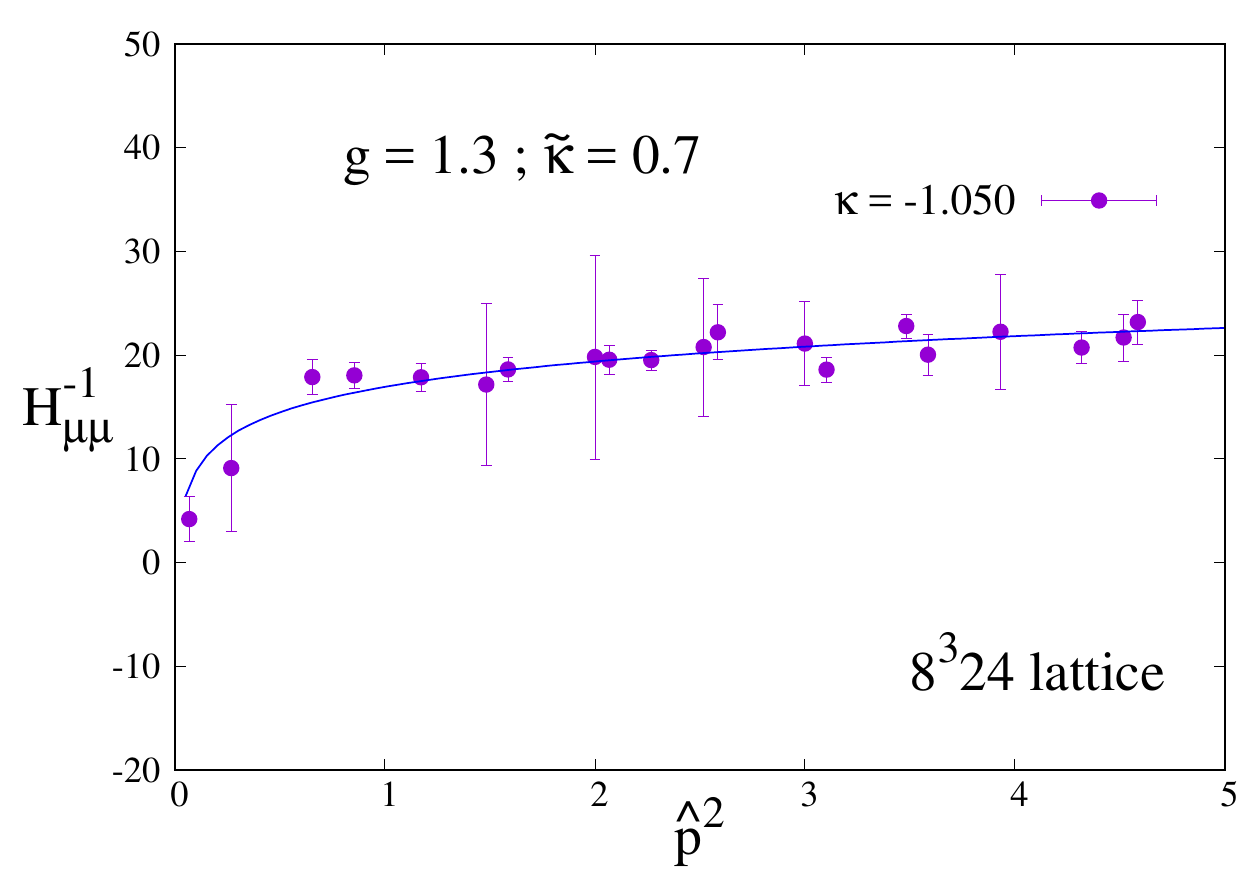}
  \end{center}
  \caption{The inverse of the effective scalar propagator versus $\hat{p}^2$ at $g=1.3$ near the continuous FM-FMD transition on $8^324$ lattices indicating decoupling of \emph{lgdof}.}
  \label{H-prop}
 \end{figure}

The scalar propagator was not investigated in \cite{DeSarkar2016}. With the expression given in Eq.~\ref{scalar-prop}, in Fig.~\ref{H-prop}, we plot its inverse with $\mu=\nu$ against $\hat{p}^2 $ at $g=1.3$ in the FM phase very close to the continuous part of the FM-FMD transition. The scalar propagator is noisy, consistent with observations made in \cite{Bock_etal2000}, despite having about 50000 equilibrated field configurations, far more than our usual number. However, the non-linearity of the inverse propagator at small momenta suggests absence of a pole. The non-linearity in the inverse propagator at small momenta was also observed at weak coupling studies in \cite{Bock_etal2000}, both through WCPT and numerical studies and was accepted as an indication of the decoupling of the \emph{lgdof}. The smooth curve in Fig.~\ref{H-prop} is essentially to guide the eye, and not a fit. However, we have observed that our inverse propagator data for small lattice momentum $\hat{p}^2$ are consistent with a non-linear behaviour like $(\log \, \hat{p}^2)^{-1}$, as found perturbatively in \cite{Bock_etal2000}.
 
\begin{figure}
 \begin{center}
  \includegraphics[width=0.48\linewidth]{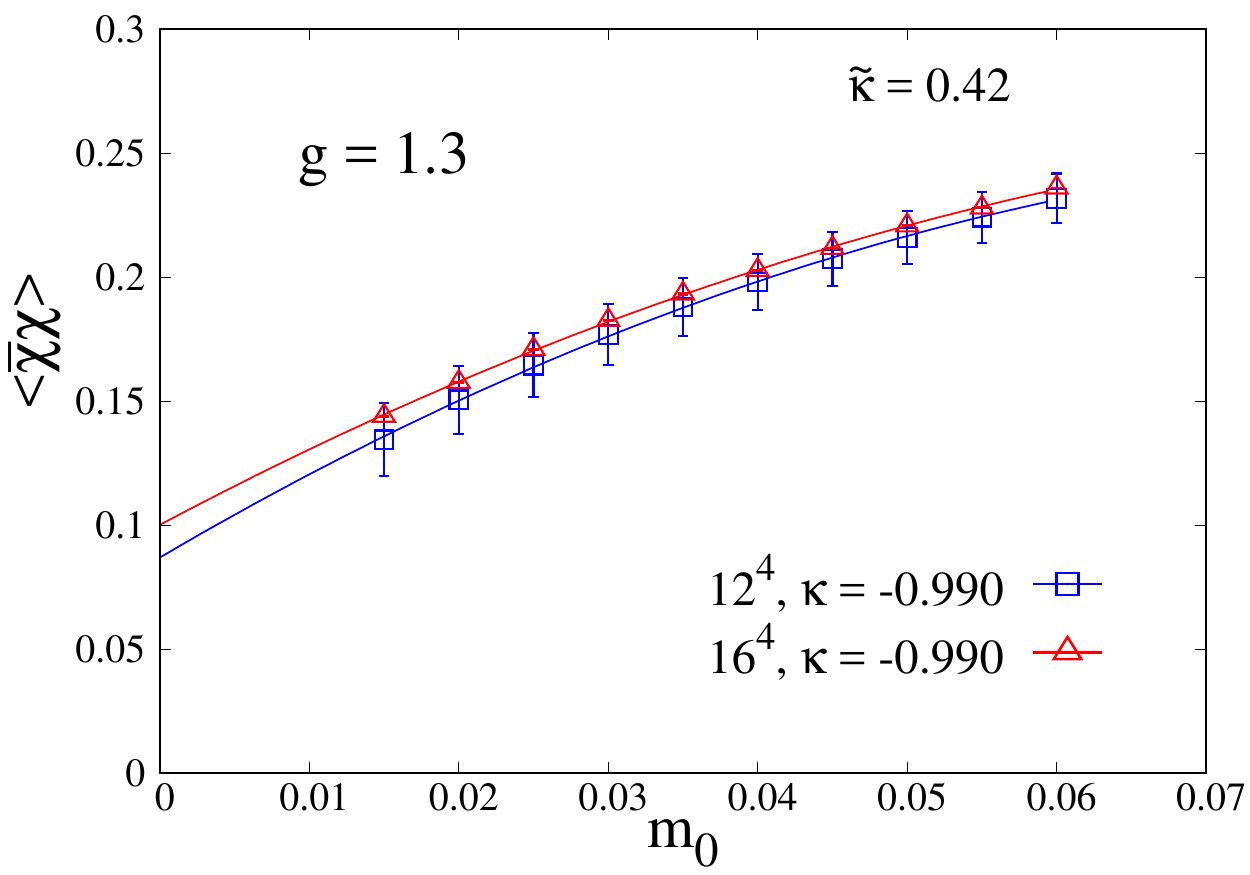} \quad
  \includegraphics[width=0.48\linewidth]{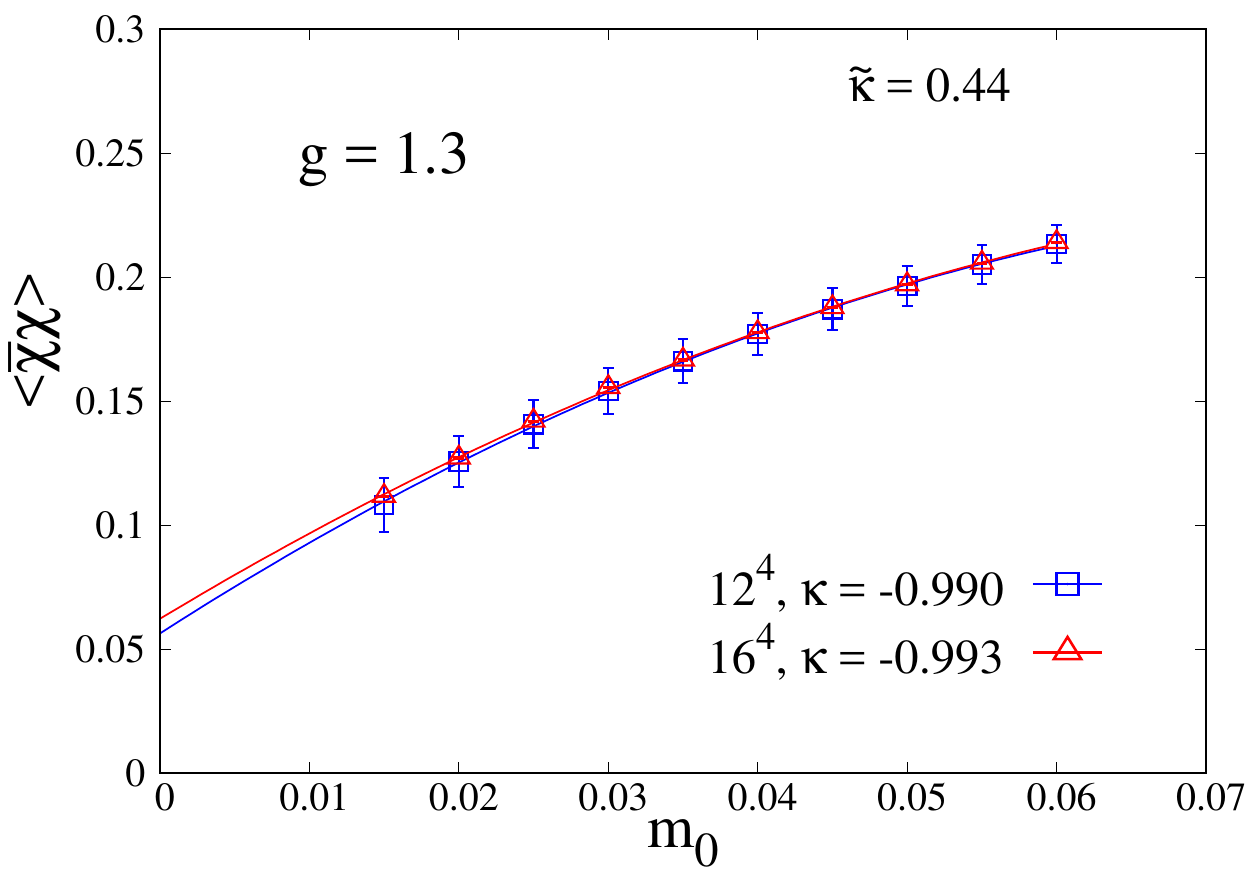} \\
  \includegraphics[width=0.48\linewidth]{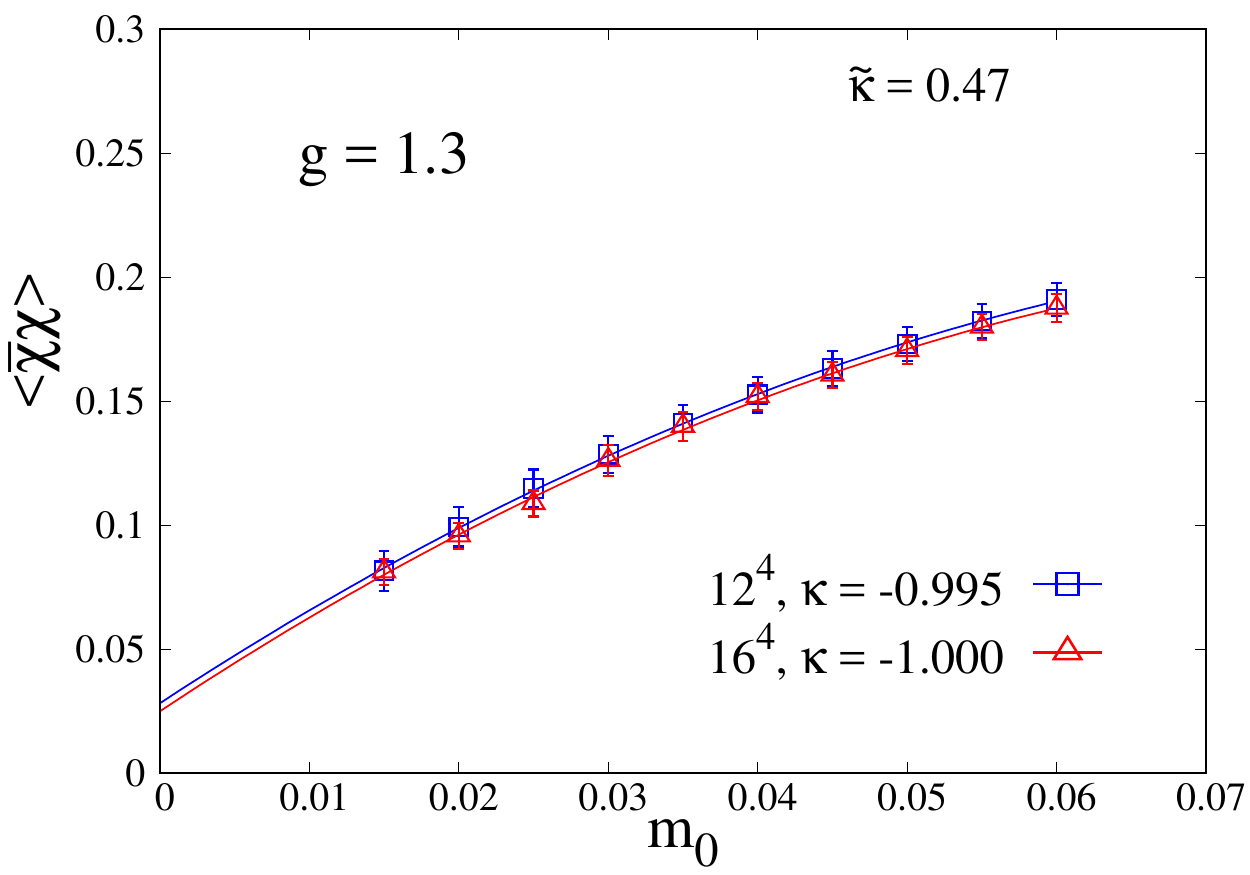} \quad
  \includegraphics[width=0.48\linewidth]{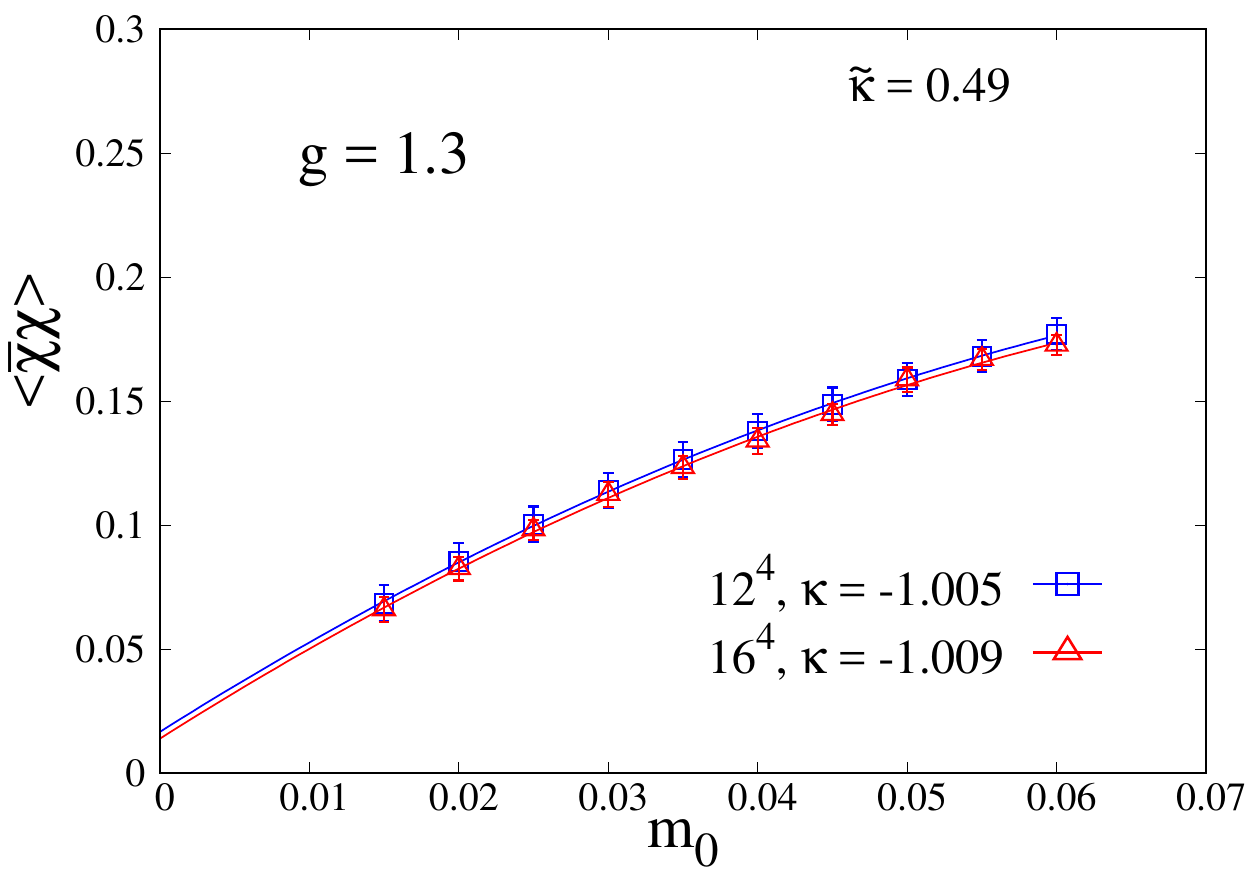}
  \end{center}
  \caption{Quenched chiral condensates versus bare fermion mass at four values of $\kt$ around the tricritical point at $g=1.3$ on two lattice volumes $12^4$ and $16^4$.}
  \label{chicond}
 \end{figure}

 In \cite{DeSarkar2016} at $g=1.3$, a chiral transition, when probed with quenched Kogut-Susskind fermions, was observed roughly around the tricritical point. While the tricritical point was found to be around $\kt=0.45$ and $\kap=-1.000$, the chiral transition around the same values of $\kap$ near the FM-FMD transition was determined to be between $\kt=0.40$ and 0.50. In this paper, we have probed the chiral transition with more precision, and our results are summarised in the four plots of Fig.~\ref{chicond}. While we observe that the chiral condensate gradually dips towards zero as $\kt$ increases, the volume dependence of the plots, especially at lower fermion masses, are very different for the lower two $\kt$ values as opposed to the higher $\kt$ values. In Fig.~\ref{chicond}, for $\kt\leq 0.44$, the $16^4$ data and their chiral extrapolation always lie above that of the $12^4$ lattice, while the trend appears to be opposite for data at $\kt \geq 0.47$. The opposite trend of volume dependence in the chiral limit is, however, very clearly seen in the corresponding plots in \cite{DeSarkar2016} at $\kt=0.40$ and 0.50. From the numerical evidence, it appears that the chiral transition takes place very near, if not coincident with, the tricritical point where the order of the FM-FMD transition changes from first order to continuous (second order). The vanishing chiral condensate at the continuous part of the FM-FMD transition is taken as an evidence of absence of non-trivial physics from this transition, although the chiral condensates do not exactly vanish on our finite lattices. 

\begin{figure}
 \begin{center}
  \includegraphics[width=0.66\linewidth]{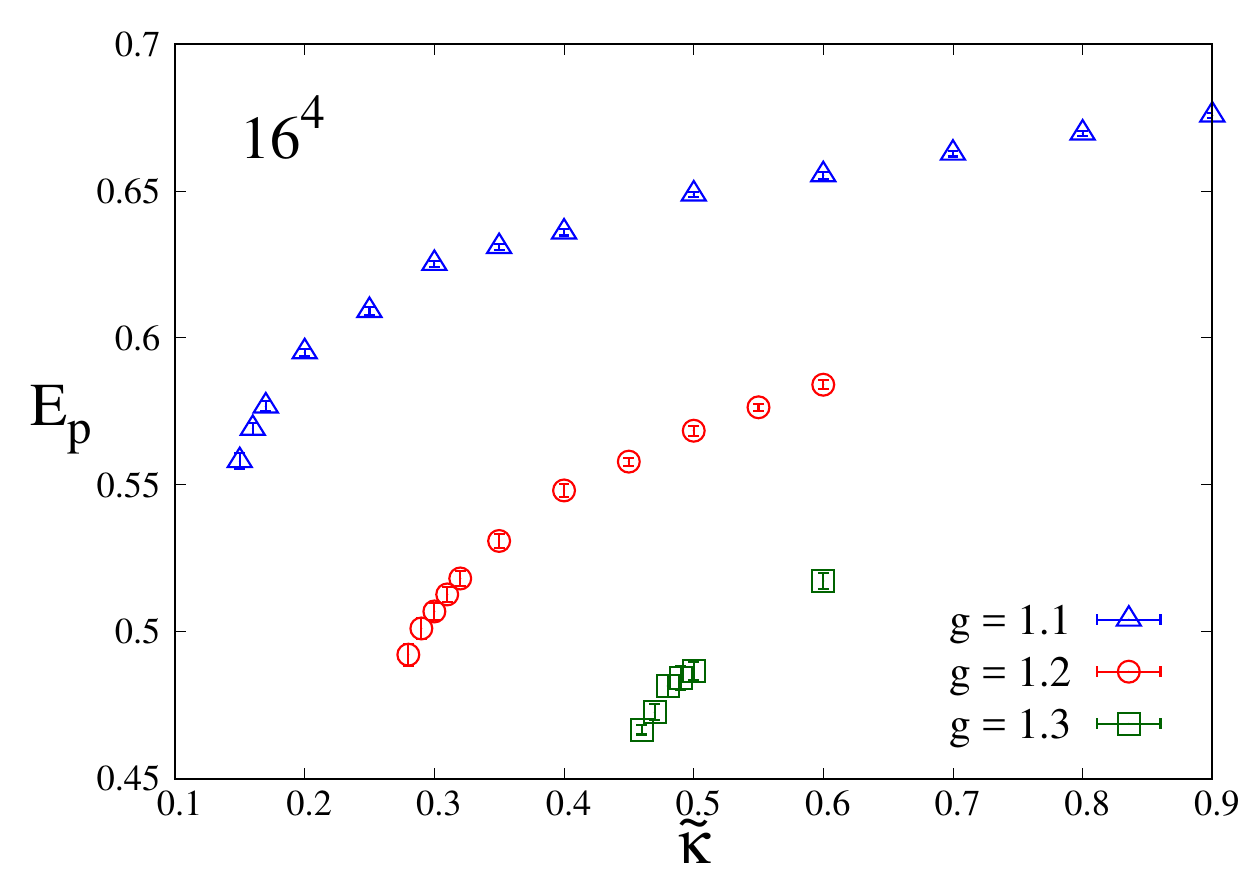}
  \end{center}
  \caption{Average plaquette near the continuous FM-FMD transition versus $\kt$ at gauge couplings $g=1.1$, 1.2 and 1.3 on $16^4$ lattice.}
  \label{plaquette}
 \end{figure}

All our numerical investigations at strong gauge couplings indicate that, given any bare gauge coupling, there always exists a continuous FM-FMD transition for a sufficiently large $\kt$ and the emerging physics while approaching the transition from the FM-side is governed by the WCPT point at $g=0$ and $\kt \rightarrow \infty$. Fig.~\ref{plaquette} shows the average plaquette in the FM phase near the continuous FM-FMD transition increasing with increasing $\kt$ for gauge couplings $g = 1.1$, 1.2 and 1.3. It is reasonable to expect that as $\kt$ is increased, the average plaquette eventually would approach unity, the value of the plaquette at the perturbative point, in a behaviour similar to that of the slope of the gauge field propagator.   

\section{Conclusion} \label{Conclusion}
A non-perturbatively gauge-fixed compact $U(1)$ lattice gauge theory is an alternate formulation of the pure $U(1)$
gauge theory on lattice. It is not only important because it provides a continuum limit (unlike the standard Wilson formulation) and a possible probe at short distance behaviour of a perturbatively non-asymptotically free theory (for example, by examining the universality class of the tricritical line obtained by us), but particularly for a manifestly local lattice formulation of abelian chiral gauge theory with lattice fermions which explicitly break chiral symmetry. It obviously is very important to know the phase diagram of the theory for wide range of all its parameters so that all possible continuum limits and the universality classes can be traced. 

We have carried out an extensive numerical investigation of the theory, especially at strong gauge couplings ($g>1$). In a previous study \cite{DeSarkar2016}, results were presented for a single gauge coupling $g=1.3$. The approach in this paper is to scan a wide range of the 3-dimensional parameter space, generating gauge field configurations for a very large number of points in that parameter space to locate and determine the nature of the phase transitions and come up with an overall picture.

We find that there is no lack of continuity between the FM-FMD phase transition near the perturbative point at $g=0$ and $\kt\rightarrow \infty$ and the FM-FMD transition at strong gauge couplings up to the edge of the tricritical line. The continuous part of the FM-FMD transition surface (blue surface, marked I in Fig.~\ref{3dPD}) is one continuous surface, and the results of all our measurements help build the emergence of a single universality class, obtained by approaching the transition near $g=0$, $\kt\rightarrow \infty$ from the FM side. Hence irrespective of the bare gauge coupling being weak or strong, at a strong enough coefficient $\kt$ of the HD gauge-fixing term, the physics obtained by approaching the continuous FM-FMD transition from the FM side is governed by the perturbative point and is a Lorentz covariant theory of free massless photons, with the redundant \emph{lgdof} decoupling at that transition.      

The tricritical line at strong gauge couplings is potentially the only place where a different universality class with non-trivial physics may appear. However, a detailed investigation in that direction deserves a dedicated study and is outside the scope of the current work, given its vast and extensive nature.

The action with the HD gauge-fixing term poses its own problems in the Monte Carlo importance sampling. We found that a local algorithm like MM is poor in generating gauge field configurations (corresponding to quantum fluctuations around the global minimum of the classical action), especially at large values of the coefficient $\kt$ of the HD term and at relatively larger lattices. A global algorithm like HMC was generally found to produce faithful field configurations and was used to generate the ensembles at the vast number of points in the 3-dimensional parameter space.

\acknowledgments
The work is supported by the DAE, Govt. of India. The computing facility of the Theory Division, SINP acquired and maintained during the 11th plan and the 12th plan periods through projects under the DAE, Govt. of India, is highly acknowledged. Authors are grateful to Richard Chang for help with system maintenance. One of the authors (MS) would like to thank A.~Ghosh and P.~K.~Mohanty for fruitful discussions on phase transitions in statistical systems.


\begin{thebibliography}{99}

\bibitem{DeSarkar2016}
A.~K.~De and M.~Sarkar,
\emph{Tricritical points in a compact $U(1)$ lattice gauge theory at strong coupling},
\emph{Phys.\ Rev.\ D} {\bf 93}, no. 11 (2016) 114504.

\bibitem{Neub87}
H.~Neuberger,
\emph{Nonperturbative {BRS} Invariance and the Gribov Problem},
\emph{Phys.\ Lett.\ B} {\bf 183} (1987) 337.  
  
\bibitem{Testa98}
M.~Testa,
\emph{Lattice gauge fixing, Gribov copies and BRST symmetry},
\emph{Phys.\ Lett.\ B} {\bf 429} (1998) 349.

\bibitem{Schad99}
M.~Schaden,
\emph{Equivariant gauge fixing of SU(2) lattice gauge theory},
\emph{Phys.\ Rev.\ D} {\bf 59} (1999) 014508. 

\bibitem{GoltSham2004}
M.~Golterman and Y.~Shamir,
\emph{SU(N) chiral gauge theories on the lattice},
\emph{Phys.\ Rev.\ D} {\bf 70} (2004) 094506.
  
\bibitem{GoltSham97}
M.~F.~L.~Golterman and Y.~Shamir,
\emph{A Gauge fixing action for lattice gauge theories},
\emph{Phys.\ Lett.\ B} {\bf 399} (1997) 148.
  
\bibitem{Sham98}
Y.~Shamir,
\emph{The Standard model from a new phase transition on the lattice},
\emph{Phys.\ Rev.\ D} {\bf 57} (1998) 132.

\bibitem{Wilson74}
Kenneth G. Wilson, 
\emph{Confinement of Quarks}, 
\emph{Phys. Rev. D} {\bf 10} (1974) 2445.

\bibitem{KarstenSmit81}
L.~H.~Karsten and J.~Smit,
\emph{Lattice Fermions: Species Doubling, Chiral Invariance, and the Triangle Anomaly},
\emph{Nucl.\ Phys.\ B} {\bf 183}, 103 (1981).

\bibitem{NN81a}
H.~B.~Nielsen and M.~Ninomiya,
\emph{Absence of Neutrinos on a Lattice. 1. Proof by Homotopy Theory},
\emph{Nucl.\ Phys.\ B} {\bf 185}, 20 (1981).

\bibitem{NN81b}
H.~B.~Nielsen and M.~Ninomiya,
\emph{No Go Theorem for Regularizing Chiral Fermions},
\emph{Phys.\ Lett.\ B} {\bf 105}, 219 (1981).

\bibitem{GinspargWilson82}
P.~H.~Ginsparg and K.~G.~Wilson,
\emph{A Remnant of Chiral Symmetry on the Lattice},
\emph{Phys.\ Rev.\ D} {\bf 25}, 2649 (1982).

\bibitem{Luscher98}
M.~Luscher,
\emph{Abelian chiral gauge theories on the lattice with exact gauge invariance},
\emph{Nucl.\ Phys.\ B} {\bf 549} (1999) 295.
  
\bibitem{BockDeSmit91}
W.~Bock, A.~K.~De and J.~Smit,
\emph{Fermion masses at strong Wilson-Yukawa coupling in the symmetric phase},
\emph{Nucl.\ Phys.\ B} {\bf 388} (1992) 243.

\bibitem{GoltPetchSmit91}
M.~F.~L.~Golterman, D.~N.~Petcher and J.~Smit,
\emph{Fermion interactions in models with strong Wilson-Yukawa couplings},
\emph{Nucl.\ Phys.\ B} {\bf 370} (1992) 51.

\bibitem{Smit}
J.~Smit,
\emph{Fermions on a Lattice},
\emph{Acta Phys.\ Polon.\ B} {\bf 17} (1986) 531.

\bibitem{Swift}
P.~V.~D.~Swift,
\emph{The Electroweak Theory on the Lattice},
\emph{Phys.\ Lett.}\  {\bf 145B} (1984) 256.

\bibitem{EP}
E.~Eichten and J.~Preskill,
\emph{Chiral Gauge Theories on the Lattice},
\emph{Nucl.\ Phys.\ B} {\bf 268} (1986) 179.

\bibitem{DWW}
D.~B.~Kaplan,
\emph{Chiral fermions on the lattice},
\emph{Nucl.\ Phys.\ Proc.\ Suppl.}\  {\bf 30} (1993) 597.
  
\bibitem{BasakDeSinha2004}
S.~Basak, A.~K.~De and T.~Sinha,
\emph{On the continuum limit of gauge fixed compact U(1) lattice gauge theory},
\emph{Phys.\ Lett.\ B} {\bf 580} (2004) 209.

\bibitem{Bock_etal2000}
W.~Bock, K.~C.~Leung, M.~F.~L.~Golterman and Y.~Shamir,
\emph{The Phase diagram and spectrum of gauge fixed Abelian lattice gauge theory},
\emph{Phys.\ Rev.\ D} {\bf 62} (2000) 034507.
 
\bibitem{Bock_etal1998}
W.~Bock, M.~F.~L.~Golterman and Y.~Shamir,
\emph{Lattice chiral fermions through gauge fixing},
\emph{Phys.\ Rev.\ Lett.}  {\bf 80} (1998) 3444.

\bibitem{Bock_etal1998_2} 
W.~Bock, M.~F.~L.Golterman and Y.~Shamir,
\emph{Chiral fermions on the lattice through gauge fixing: Perturbation theory},
\emph{Phys.\ Rev.\ D} {\bf 58} (1998) 034501.

\bibitem{BasakDe2001}
S.~Basak and A.~K.~De,
\emph{Chiral gauge theory on lattice with domain wall fermions},
\emph{Phys.\ Rev.\ D} {\bf 64} (2001) 014504.

\bibitem{BasakDe2001_2}
S.~Basak and A.~K.~De,
\emph{Gauge fixed domain wall fermions on lattice at small Yukawa coupling},
\emph{Phys.\ Lett.\ B} {\bf 522} (2001) 350. 

\bibitem{Bock1997fu}
W.~Bock, M.~F.~L.~Golterman and Y.~Shamir,
\emph{On the phase diagram of a lattice U(1) gauge theory with gauge fixing},
\emph{Phys.\ Rev.\ D} {\bf 58} (1998) 054506.
 
\bibitem{Borelli_etal1990}
A.~Borrelli, L.~Maiani, R.~Sisto, G.~C.~Rossi and M.~Testa,
\emph{Neutrinos on the Lattice: The Regularization of a Chiral Gauge Theory},
\emph{Nucl.\ Phys.\ B} {\bf 333} (1990) 335.
   
\bibitem{U1late1}
G.~Arnold, B.~Bunk, T.~Lippert and K.~Schilling,
\emph{Compact QED under scrutiny: It's first order},
\emph{Nucl.\ Phys.\ Proc.\ Suppl.}  {\bf 119}, (2003) 864.

\bibitem{U1late2}
M.~Vettorazzo and P.~de Forcrand,
\emph{Electromagnetic fluxes, monopoles, and the order of the 4-d compact U(1) phase transition},
\emph{Nucl.\ Phys.\ B} {\bf 686} (2004) 85.

\bibitem{U1late3}
M. Panero,
\emph{A Numerical study of confinement in compact QED},
\emph{JHEP} 0505 (2005) 066.

\bibitem{Evertz_etal1987}
H.~G.~Evertz, K.~Jansen, J.~Jersak, C.~B.~Lang and T.~Neuhaus,
\emph{Photon and Bosonium Masses in Scalar Lattice QED},
\emph{Nucl.\ Phys.\ B {\bf 85} [FS19] (1987) 590}.

\bibitem{Bitar_etal1989}
K.~Bitar, A.~D.~Kennedy, R.~Horsley, S.~Meyer, P.~Rossi,
\emph{The QCD Finite Temperature Transition and Hybrid Monte Carlo},
\emph{Nucl.\ Phys.\ B {\bf 313} (1989) 348}.

\bibitem{Somen}
R.~Kapri and S.~M.~Bhattacharjee, 
\emph{Manipulating a single adsorbed DNA for a critical endpoint},
\emph{Euro.\ Phys.\ Lett.}\, 83 (2008) 68002.

\end{thebibliography}
\end{document}